\documentclass[journal]{IEEEtran}

\usepackage{ifpdf}

\usepackage{cite}

\ifCLASSINFOpdf
   \usepackage[pdftex]{graphicx}
   \graphicspath{{fig/}}
\else
\fi
\usepackage{amsmath}

\ifCLASSOPTIONcompsoc
  \usepackage[caption=false,font=normalsize,labelfont=sf,textfont=sf]{subfig}
\else
  \usepackage[caption=false,font=footnotesize]{subfig}
\fi

\usepackage{float}
\usepackage{graphicx,color}
\usepackage{siunitx}
\usepackage{url}
\usepackage{bm}
\usepackage{eucal}
\usepackage{dsfont}
\usepackage{amsmath,amssymb,amsfonts}
\usepackage{algorithmic}
\usepackage{import}
\usepackage{flushend}
\usepackage[hidelinks]{hyperref}

\usepackage{nomencl}

\DeclareMathOperator{\E}{\mathbb{E}}

\begin{document}
%
\title{An Interpretable Probabilistic Model for Short-Term Solar Power Forecasting Using Natural Gradient Boosting}
%
%
%

\author{Georgios~Mitrentsis
        and~Hendrik~Lens 
\thanks{G. Mitrentsis and H. Lens are with the Institute of Combustion and Power Plant Technology, University of Stuttgart, Stuttgart 70569, Germany (e-mail: georgios.mitrentsis; hendrik.lens@ifk.uni-stuttgart.de)}
}

\maketitle

\begin{abstract}
PV power forecasting models are predominantly based on machine learning algorithms which do not provide any insight into or explanation about their predictions (black boxes). Therefore, their direct implementation in environments where transparency is required, and the trust associated with their predictions may be questioned. To this end, we propose a two stage probabilistic forecasting framework able to generate highly accurate, reliable, and sharp forecasts yet offering full transparency on both the point forecasts and the prediction intervals (PIs). In the first stage, we exploit natural gradient boosting (NGBoost) for yielding probabilistic forecasts, while in the second stage, we calculate the Shapley additive explanation (SHAP) values in order to fully comprehend why a prediction was made. To highlight the performance and the applicability of the proposed framework, real data from two PV parks located in Southern Germany are employed. Comparative results with two state-of-the-art algorithms, namely Gaussian process and lower upper bound estimation, manifest a significant increase in the point forecast accuracy and in the overall probabilistic performance. Most importantly, a detailed analysis of the model's complex nonlinear relationships and interaction effects between the various features is presented. This allows interpreting the model, identifying some learned physical properties, explaining individual predictions, reducing the computational requirements for the training without jeopardizing the model accuracy, detecting possible bugs, and gaining trust in the model. Finally, we conclude that the model was able to develop complex nonlinear relationships which follow known physical properties as well as human logic and intuition.
\end{abstract}   


%
\IEEEpeerreviewmaketitle

\makenomenclature
\nomenclature{$ \bm{x} $}{Feature vector}
\nomenclature{$ y $}{Target Variable}
\nomenclature{$ \theta $}{Parameters of a probability distribution}
\nomenclature{$ P $}{Probability distribution}
\nomenclature{$ f^{(n)} $}{Base learners of stage $n$}
\nomenclature{$ \CMcal{L} $}{Scoring rule}
\nomenclature{$ \CMcal{T} $}{Training set}
\nomenclature{$ M $}{Total number of training examples}
\nomenclature{$ i $}{Training example index}
\nomenclature{$ g^{(n)} $}{Natural gradient at stage $n$}
\nomenclature{$ \CMcal{I} $}{Fisher information}
\nomenclature{$ \E $}{Expected value}
\nomenclature{$ \rho^{(n)} $}{Scaling factor at stage $n$}
\nomenclature{$ \mu $}{Mean value}
\nomenclature{$ \sigma $}{Standard deviation}
\nomenclature{$ \CMcal{N} $}{Gaussian distribution}
\nomenclature{$ \bm{K} $}{Covariance matrix}
\nomenclature{$ C $}{Kernel function}
\nomenclature{$ l, \alpha, p $}{The lengthscale, the scale-mixture, and the period of a kernel}
\nomenclature{$ \epsilon $}{Gaussian noise}
\nomenclature{$ \bm{\Sigma} $}{Covariance matrix of the noise-free system}
\nomenclature{$ \bm{I} $}{Identity matrix}
\nomenclature{$ \bm{k}$}{Covariance vector}
\nomenclature{$ L $}{Lower bound of predictions intervals}
\nomenclature{$ U $}{Upper bound of predictions intervals}
\nomenclature{$ R $}{Maximum value of the target variable}
\nomenclature{$ c $}{Binary coefficient if a predicted value is within the prediction intervals}
\nomenclature{$ \eta $}{Penalty factor of lower upper bound estimation algorithm }
\nomenclature{$ \gamma $}{Hyperparameter of lower upper bound estimation algorithm }
\nomenclature{$ \hat{F} $}{Predicted cumulative distribution function}
\nomenclature{$ \mathds{1} $}{Heaviside function}
\nomenclature{$ g $}{Explanation model (SHAP method)}
\nomenclature{$ \phi $}{SHAP value}
\nomenclature{$ z^{\prime} $}{Binary coefficient if a feature is present or not}
\nomenclature{$ d $}{Observed feature}
\nomenclature{$ h $}{Function mapping a binary vector into the original input space}
\nomenclature{$ \CMcal{Z} $}{Set containing all features}
\nomenclature{$ \CMcal{S} $}{Subset of features}
\nomenclature{$ \Phi $}{SHAP interaction value}

\printnomenclature

\section{Introduction}
%
%
%
%
\IEEEPARstart{T}{he} ongoing transition of the power system towards a fossil-fuel-free system has led to a wide integration of renewable energy sources (RES) worldwide. Photovoltaic (PV) power forms one of the most widespread and promising alternatives to conventional power plants \cite{ahmed2020review}. However, the stochastic nature of PV power induced by volatile weather conditions hinders the reliable electricity supply, which can result in an increase in the reserve capacity of the system. At the same time, PV parks are typically connected to the grid through power electronics leading to declining system inertia and thus, pushing the system closer to its stability margins \cite{hatziargyriou2020stability}. Accurate and reliable PV power forecasting can alleviate the aforementioned challenges allowing for a large-scale PV integration. 

PV power forecasting can be split into two general categories, namely deterministic and probabilistic approaches. The former yield single point forecasts at each time step whereas in the latter, the PV output is considered as a random variable and the model generates the respective probability density functions or prediction intervals (PIs) \cite{van2018probabilistic}. While deterministic forecasting has been extensively studied in the literature, probabilistic approaches have only recently gained attention \cite{wang2019review}. For a detailed overview of the current deterministic approaches, we refer to \cite{das2018forecasting,wang2019review,ahmed2020review}.

Probabilistic forecasts are more valuable compared to deterministic ones since they provide uncertainty information about the future power outcomes \cite{wang2019review}. This information can be leveraged by the involved parties, e.g., TSOs, DSOs, traders, legislators, for a lower risk and a more beneficial decision making \cite{van2018probabilistic}. Indicative examples can be found in \cite{doherty2005new} and \cite{papavasiliou2011reserve}, which exploit probabilistic load and wind power forecasts in order to estimate the optimal reserve requirements. Furthermore, trading wind power in the day-ahead market using probabilistic predictions has shown an increase in the profit of the producer in comparison with the case of traditional deterministic predictions, as presented in \cite{pinson2009ensemble} and \cite{alessandrini2014comparison}.

The probabilistic forecasting methods can be further split into parametric and nonparametric approaches depending on the output distribution assumptions \cite{van2018review}. Parametric approaches assume that the PV power follows a known probability distribution, the parameters of which should be estimated during the training procedure. In \cite{bracale2013bayesian}, a Gamma distribution is deployed and its parameters are estimated using Bayesian inference. Similarly, a Gaussian distribution with zero mean and a variable standard deviation is selected in \cite{lorenz2009irradiance} in order to describe the irradiance errors. The Guassian distribution was also employed in \cite{fonseca2015use}, where it is compared with the Laplacian distribution with respect to modeling the forecast errors. Generally, most of the existing parametric approaches focus on extending their deterministic predictions into probabilistic ones. As a result, nonparametric methods, which make no assumptions about the output distribution, have dominated the literature \cite{wang2019review}. 

The nonparametric approaches can be grouped into six different categories based on the main method that is being employed. Those include quantile regression, bootstrapping, lower upper bound estimation (LUBE), gradient boosting, kernel density estimation, and analog ensemble \cite{wang2019review}. A major part of the state-of-the-art approaches is summarized below.

\textit{Quantile regression}: Quantile regression can be seen as an extension of linear regression for probabilistic output. In \cite{lauret2017probabilistic}, three different quantile regression models using different sets of features are developed and analyzed, while \cite{juban2016multiple} presents a multiple quantile regression method in order to estimate the full probability distributions. References \cite{van2018probabilistic} and \cite{wen2019performance} study the application of quantile regression in PV power forecasting and compare it with a Gaussian process (GP) and a bootstrap model, respectively. 

\textit{Bootstrapping}: Bootstrapping is a simple statistical method for yielding PIs, which has been widely deployed in probabilistic forecasting \cite{wang2019review,ahmed2020review}. Three representative examples of this method in PV power forecasting can be found in \cite{wen2019performance,alhakeem2015new,grantham2016nonparametric}. In addition, \cite{bozorg2021bayesian} introduces the application of Bayesian bootstrapping in real-time probabilistic PV forecasting, while \cite{li2018interval} proposes an improved bootstrap method that requires less assumptions about the distribution of the forecast errors.

\textit{LUBE}: A recently introduced probabilistic regression algorithm based on neural networks with applications in various fields is the lower upper bound estimation (LUBE) \cite{khosravi2010lower}. This algorithm was combined with extreme learning machines in \cite{ni2017ensemble}, in order to generate accurate PIs, while a new model initialization approach was proposed in \cite{li2019solar}. In \cite{pan2020probabilistic}, a stacked auto-encoder is used as a pre-processing step to reduce the inputs of the LUBE algorithm. A similar approach combining extreme learning machines and an auto-encoder was introduced in \cite{long2021combination}.

\textit{Gradient boosting}: Gradient boosting is based on the principle of combining many weak learners in order to develop a powerful probabilistic machine learning model. In \cite{huang2016semi}, a gradient boosting regression tree model is utilized to develop point forecasts while a $k$-nearest neighbors regression model is adopted for estimating the respective PIs. In \cite{verbois2018probabilistic}, a principal component analysis is combined with a quantile gradient boosting algorithm for the day-ahead solar irradiance forecasting. Lastly, feature engineering techniques are incorporated with gradient boosting trees in order to improve the accuracy of a numerical weather prediction model in \cite{andrade2017improving}.

\textit{Kernel density estimation}: Kernel density estimation methods are considered a popular choice for RES power forecasting due to their flexibility and adaptability \cite{zhang2015probabilistic}. Applications of kernel density estimation methods in PV power forecasting have been presented in \cite{chai2016nonparametric,yamazaki2015improvement,liu2018prediction, lotfi2020novel,pan2021prediction}.
 
\textit{Analog ensemble}: The analog ensemble is a hybrid model that employs numerical weather predictions and past power values in order to yield probabilistic forecasts. The algorithm was initially proposed in \cite{alessandrini2015analog} and was further developed by adding a neural network component in \cite{cervone2017short}. Another application of the analog ensemble algorithm in PV power forecasting can be found in \cite{alessandrini2020schaake}.

There are a few more works that do not belong to one of the aforementioned categories. For instance, in the recent work of \cite{sun2020probabilistic}, the authors developed a weather scenario generation method for yielding various weather scenarios, which are then being applied on the generation of probabilistic forecasts. In \cite{voyant2014bayesian}, a methodology combining Bayesian rules and stochastic models was introduced, while a dynamic Gaussian process (GP) model was proposed in \cite{van2018probabilistic2}. Finally, the output of a numerical weather prediction ensemble for solar power forecasting is post-processed using a Bayesian model averaging in the work of \cite{doubleday2020probabilistic}.

Yet, despite the inspiring research work in probabilistic RES forecasting, it has still not been studied to a great extent. At the same time, we also lack a systematic approach to integrate it into the system operation \cite{wang2019review,sun2020probabilistic}. Furthermore, apart probably from the bootstrapping techniques, those machine learning approaches can be considered as complex black box models that require high computational resources for training. Those characteristics may hinder the widespread integration of those algorithms in practice, especially in the case of TSOs and DSOs who are responsible for the safe and reliable operation of the power system. The more complex become those models, the harder it gets to understand them and explain their results. Therefore, in applications where safety critical decisions are to be made, e.g., power system operation and control, the trust in complex black box models may be questioned. At the same time, due to the role of TSOs and DSOs in society, one of their main duties is to be transparent and share information about the forecasting quantities with the general public. Hence, their trust in those models is of utmost importance. It is also worth mentioning that the forecasting models are broadly employed by energy traders. Therefore, identifying learned patterns or rules, detecting possible bias in the predictions, and exploring regions where the used features are not sufficient for an accurate prediction may be crucial for the developed bidding strategies and consequently, for the maximization of their profit. Moreover, debugging a machine learning models is a rather challenging task. Even at big technology companies, many bugs in machine learning pipelines may not be discovered \cite{zinkevich2017rules}. Therefore, the ability to interpret the generated forecasts has been characterized as \textit{``vital"} in the very recent work of \cite{ahmed2020review}.

To this end, the aim of this work is to introduce a PV forecasting model that generates accurate, reliable, and sharp probabilistic predictions, minimizes the domain knowledge required for its practical implementation, and provides full transparency on its output. To fulfill these objectives, we propose a readily applied two stage framework that can generate high quality probabilistic forecasts while being fully interpretable. In the first stage, we propose the application of the natural gradient boosting (NGBoost) algorithm for yielding probabilistic PV power forecasts. NGBoost is a simple modular yet powerful gradient boosting algorithm, which can be used out-of-the-box and without requiring expert programming knowledge or time-consuming hyperparameter tuning \cite{duan2020ngboost}. Due to its ability to deploy decision trees as base learners, we combine it with Shapley additive explanation (SHAP) values in the second stage. The estimation of SHAP values is the unique consistent attribution method that is able to provide theoretical optimal explanations about the predictions of a model \cite{lundberg2020local}.

It is worth mentioning that initial work deploying SHAP values in PV power forecasting has been presented in \cite{kuzlu2020gaining} and \cite{lu2021neural}. However, both studies aim at interpreting deterministic models and do not consider SHAP interaction values. In particular, these interaction values form one of the most important points in model interpretability, since they reveal complex interactions between pairs of features.

Applications of machine learning algorithms combined with SHAP explanations have been recently introduced in the literature for a wide range of use cases. Those include building ventilation \cite{park2021comparative}, fraud detection \cite{santos2021gradient}, gold price forecasting \cite{jabeur2021forecasting}, cases of COVID-19 forecasting \cite{jing2021cross}, market sales \cite{antipov2020interpretable}, particle concentrations \cite{xu2020gradient}, load forecasting \cite{bellahsen2021aggregated}, pavement performance \cite{guo2021ensemble}, construction \cite{bakouregui2021explainable}, and traffic accident detection \cite{parsa2020toward}. However, all of these studies deploy deterministic models which do not estimate uncertainty. As a result, no further interpretation can be performed apart from deterministic predictions. In addition, these studies only briefly discuss SHAP values in order to explain the derived models, without using them to increase the model performance as in the proposed approach (Section V). Lastly, apart from \cite{jabeur2021forecasting,bakouregui2021explainable,parsa2020toward}, no other approach explores pairwise interactions between the features.

Therefore, the main contributions of this paper can be summarized in the following points:
\begin{enumerate}
\item We propose the application of a simple yet highly accurate probabilistic regression algorithm in PV power forecasting.
\item We present a thorough comparison of its performance with two state-of-the-art probabilistic algorithms while considering the influence of the season.
\item We introduce for the first time, to the best of our knowledge, the application of SHAP values in the estimation of PIs in order to understand how the algorithm models the uncertainty.
\item We calculate and analyze the SHAP interaction values for both the point forecasts and the derived PIs in order to identify whether the model has learned some known physics based relationships between the deployed features.
\item To the best of our knowledge, it the first attempt to fully interpret a probabilistic machine learning model in the context of power systems research.
\item We leverage the SHAP explanations in order to increase the model's performance by optimally selecting the most important features.
\end{enumerate}

The rest of the paper is organized as follows: In Section II, an explanatory data analysis is conducted in an attempt to identify important correlations between the data. In Section III and IV, we establish the theoretical background of the deployed forecasting algorithms, performance metrics, and interpretability methods. In Section V, we present the results of this research work, which are then discussed in Section VI. Finally, the corresponding conclusions and suggestions for future work can be found in Section VII.

\section{Power and meteorological data}

\begin{figure*}[h!] 
\centering
\includegraphics[width=0.8\textwidth]{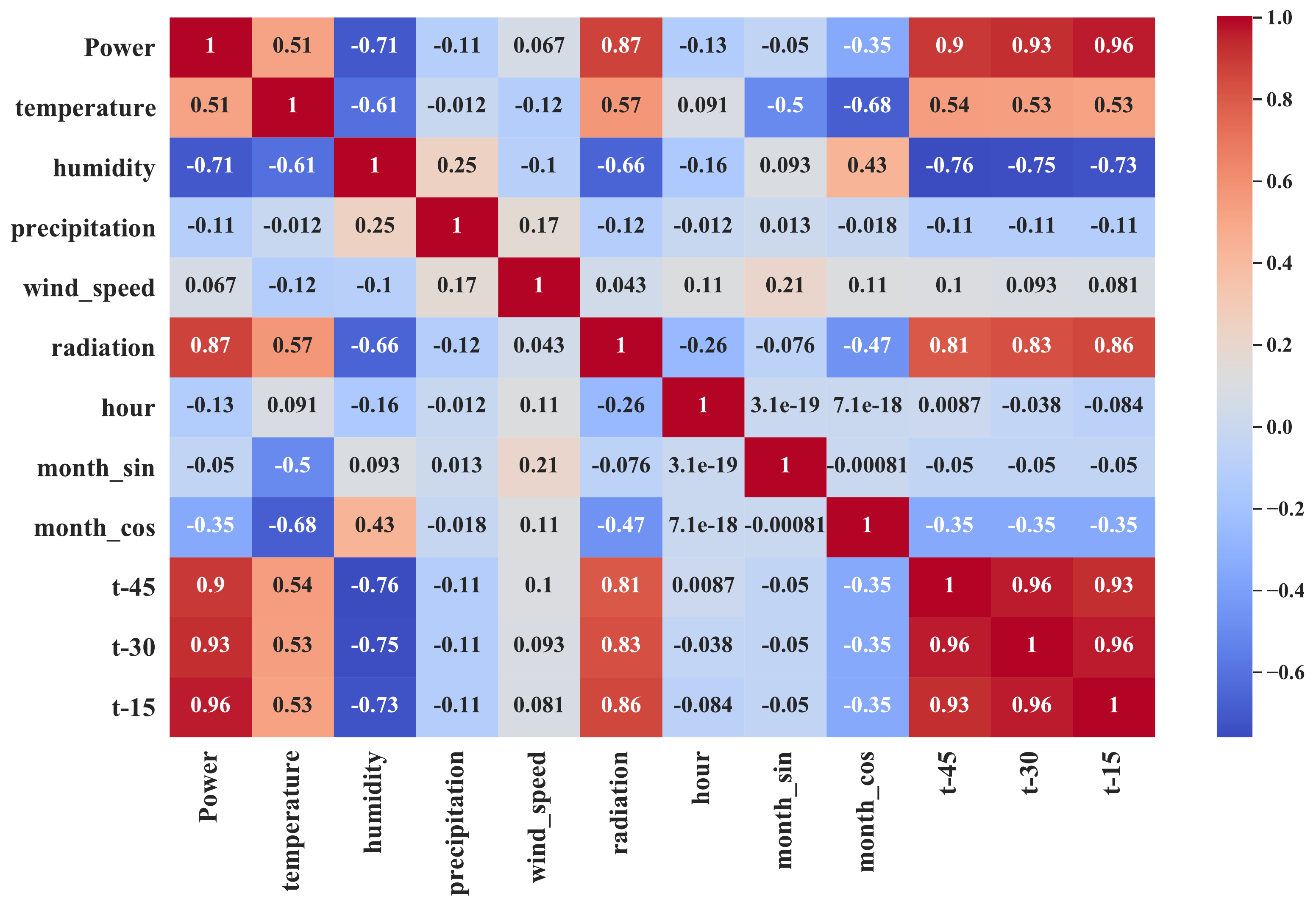}
\caption{Heatmap of Pearson correlation coefficients showing the linear correlations between all the possible pairs formed by the variables of the dataset.}
\label{fig:correlations}
\end{figure*}

In this section, we introduce the various data employed for this study. We also present a small explanatory data analysis in order to identify some pairwise relationships between the data. Those relationships will be later contrasted with the ones that the model learned in Section V. 

The power data employed in this study were acquired from two PV parks located in Southern Germany,  in particular in the state of Baden-Württemberg. They have a nominal power of approximately \SI{3.2}{MVA} and \SI{1.8}{MVA} and in the rest of the paper, we will refer to them as \textit{PVP1} and \textit{PVP2}, respectively. Both PV parks experience similar weather conditions since they are located \SI{50}{km} apart from each other. The recorded data were obtained from February 2018 to October 2019 with a time resolution of \SI{15}{min} (quarter hour data). Since there is no power generation during the night hours, the power values from 22:00 to 06:00 are discarded from the recorded dataset, as in \cite{persson2017multi}.

As we want to determine how a model handles the various features in order to make a prediction, i.e., model interpretation, we deploy a set populated by the most commonly used meteorological variables, time variables, and lagged power values as features for the model. These include temperature, relative humidity, precipitation, wind speed, ground level solar radiation, month, time of the day, and lagged power values \cite{ahmed2020review,das2018forecasting}. Based on the autocorrelation values of the power time series, we deploy three lagged power values corresponding to the power generation \SI{15}{min}, \SI{30}{min}, and \SI{45}{min} before the desired time step. We will refer to them as t-15, t-30, t-45, respectively. It is worth mentioning that the power values of the same time of the previous day (t-24h) were initially included in the model without leading to an increase in the prediction accuracy; hence, they were discarded. Furthermore, the meteorological data were obtained from the German Weather Agency (Deutscher Wetterdienst - DWD) \cite{dwd}. Those data were acquired from weather stations located a few kilometers away from the respective PV parks and thus, they do not match the exact weather values at the desired locations. In this regard, the deployed weather data are assumed as the corresponding weather forecasts at the locations of the two PV parks. We follow this assumption since we do not have weather data at the exact PV park locations and we expect only small weather variations between the locations of the weather station and the respective PV park. Under those conditions, we use these inexact weather data as an equivalent of the daily weather forecasts that we would get from a weather forecast provider for the location of the PV parks.

To ensure that December and January are seen as two consecutive months by the model and not as two completely different feature values, i.e., $\text{month}=1$ or $\text{month}=12$, we map the cyclical month variable onto a unit circle. For each month we compute its $\sin$ and $\cos$ components as:
\begin{align}
\label{eq:month_transform}
\text{month\_sin} = \sin(2 \pi \cdot \text{month}/12 ), \\
\text{month\_cos} = \cos(2 \pi \cdot \text{month}/12 ).
\end{align}
Note that, since we remove the night hours from our dataset, the time of the day is not a cyclical variable anymore and thus, no similar transformation is required.  

A potential use case of the proposed model is to support the placement of bids of specific volumes in the European short term electricity market. Therefore, we are interested in short-term forecasting for a time horizon of \SI{36}{h} (day-ahead) and a time resolution of \SI{15}{min}. To do so, we perform recursive multi-step predictions based on the predicted values of the previous time steps. In addition, we follow the sliding horizon for the model validation, where the training set comprises the data from one whole year while the test set comprises the data from the month right after the end of the training set. It is worth mentioning that similar results were obtained using the whole dataset which covers approximately 1.5 years. Nevertheless, we opted for one year data for training in order to have a balanced dataset for model interpretation. In order to thoroughly assess the performance of the algorithm under different weather and generation scenarios, four different pairs of training-test sets are constructed for each PV park, where each test set corresponds to a month of a different season.

\begin{figure}[t!] 
\centering
\includegraphics[width=\linewidth]{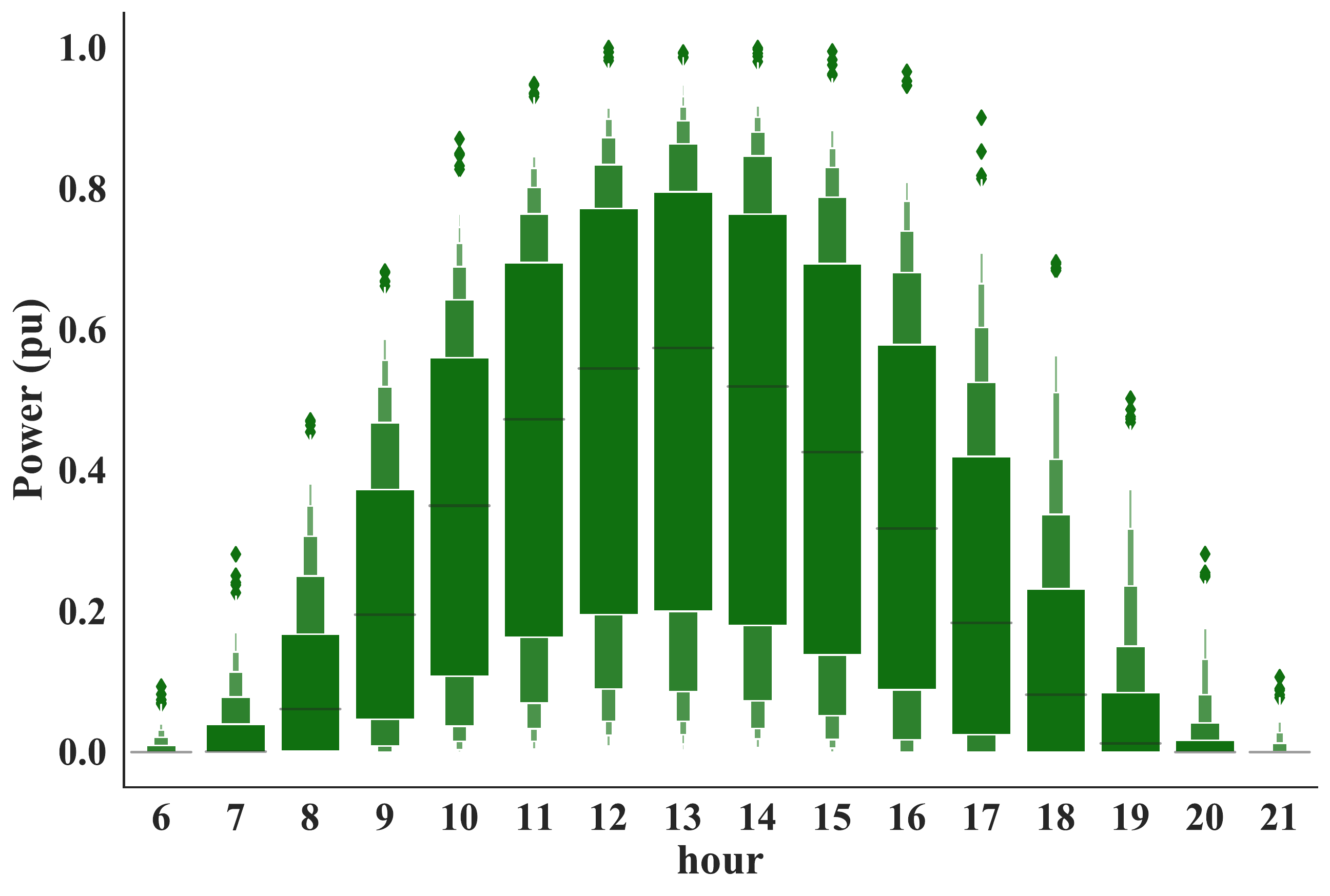}
\caption{Boxplots of output power for different time of the day. The widest box corresponds to 50\% of the data, the next two wider to 25\%, the next two wider to 12.5\%, and so on. The single dot points are considered outliers. This plot highlights how the time of the day influences the PV power generation throughout a year.}
\label{fig:power_hour}
\end{figure}

Since we are going to examine pairwise feature interactions developed by the proposed models in Section V, it is helpful to estimate the respective correlation coefficients first. Due to the fact that the correlation coefficients between the feature variables in the two PV parks were identical, we illustrate the combined correlation coefficients of PVP1 and PVP2 in Fig.~\ref{fig:correlations}. To do so, the power values were scaled using the nominal power of each park. Note that combining the two datasets was done to avoid showing identical correlation heatmaps; however, separate models were developed for each PV park. It is worth pointing out that the Pearson correlation coefficient is employed, which reveals only the linear correlation between variable pairs. Therefore, nonlinear correlations are not reflected in this correlation coefficient. As expected, the output power is strongly correlated with the power of the previous time steps as well as with the ground level radiation. Slightly less, yet still clearly correlated with the power output are the temperature and the humidity, which both have a direct impact on the efficiency of the PV panel. Small negative correlation is also observed for $\text{month\_cos}$, which can be justified by the fact that negative $\cos$ values correspond to the left half of the unit circle (spring and summer months). In addition, the time of the day (hour) is weakly correlated with the output power. This, however, does not imply that the former does not have an influence on the latter. The correlation coefficient captures only linear relationships and due to the bell-shaped power curve over one day, as shown in Fig.~\ref{fig:power_hour}, the respective correlation value is small. In this regard, the derived models will exploit this nonlinear relationship in order to make accurate predictions, as it will be revealed when interpreting the models. The rest of the pairwise relationships will help us to understand the models better in Section V. 

\section{Forecast models and metrics}
In this section, we describe the probabilistic forecasting algorithm NGBoost, which is applied in the first stage of the proposed framework. To benchmark the performance of NGBoost in PV power forecasting, we follow the recommendations of \cite{doubleday2020benchmark}, where it is advised to benchmark a forecasting model with a highly reliable but more naive model and a state-of-the-art method. Within our study, we opt for persistence as the reliable model and Gaussian process (GP) as well as lower upper bound estimation (LUBE) as two state-of-the-art methods. To do so, a set of validation metrics assessing the accuracy, bias, reliability, sharpness, and overall probabilistic performance of the forecasting models is deployed. Their scope and analytic expressions are presented in this section.

\subsection{Natural gradient boosting (NGBoost)}
NGBoost is a gradient boosting algorithm for solving probabilistic regression problems \cite{duan2020ngboost}. In general, gradient boosting algorithms are based on the sequential training of several base learners, which all together form an additive ensemble. Each learner is optimized by minimizing the current residual as estimated by the ensemble of the previous learners. Then, the output of that learner is scaled by a learning rate and it is appended to the current ensemble.   

\begin{figure}[b]
\centering
\def\svgwidth{1.0\columnwidth}
\fontsize{8pt}{11pt}\selectfont
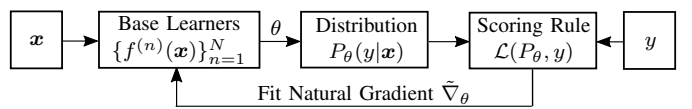
\caption{Block diagram representation of NGBoost showing its major components, i.e., base learners, distribution, scoring rule, and natural gradient \cite{duan2020ngboost}.}
\label{fig:block_ngboost}
\end{figure}

As for NGBoost, in its general form, the algorithm aims at estimating the parameters of a probability distribution $P_{\theta}(y|\bm{x})$; where $\bm{x} \in \mathbb{R}^D$ is the feature vector of an observation, $y \in \mathbb{R}$ is the target variable, and $\theta \in \mathbb{R}^L$ is the parameter vector of the distribution. For instance, if we assume a normal distribution, $\theta$ will contain the mean and the standard deviation. NGBoost is composed of three main modules: 1) the base learners $f^{(n)}$, 2) a parametric probability distribution $P_{\theta}(y|\bm{x})$, and 3) an adequate scoring rule $\CMcal{L}(P_{\theta},y)$, e.g., continuous ranked probability score (CRPS). A schematic representation of NGBoost is shown in Fig.~\ref{fig:block_ngboost}.

Regarding the base learners, shallow decision trees have been proved to be an effective choice in practice \cite{duan2020ngboost} and thus, they are also employed in this work. The next step is to determine the type of the probability distribution $P_{\theta}$. In this context, any distribution characterized by continuous parameters can be used, e.g., normal, lognormal, or exponential distributions. Preliminary results showed that the normal distribution can yield accurate and sharp forecasts and therefore, it was selected for this study. One should bear in mind that this distribution does not refer to the target variable as a whole, as in the case of parametric probabilistic forecasting approaches, but rather corresponds to the training examples that will end up at each leaf of a tree. Hence, each leaf will have its own normal distribution. Finally, the module of the scoring rule can be seen as the counterpart of the cost function in the deterministic regression. The scoring rule generates a rating given a predicted distribution and an observation of the target variable. The most commonly used scoring rule is the logarithmic score, which can mathematically described as
\begin{equation}
\CMcal{L}(\theta, y) = - \text{log}(P_{\theta}(y)).
\label{eq:mle}
\end{equation}
However, since the ``difference" between two parameters of a distribution does not reflect the ``difference" of the distributions themselves, traditional gradient based methods, e.g., gradient descent, would not work with respect to minimization of the scoring rule. To this end, NGBoost leverages the natural gradient $\tilde{\nabla}\CMcal{L}(\theta, y)$, which is motivated by information geometry, in order to learn the model parameters. This not only results in a parametrization invariant optimization but also makes the learning more stable and efficient \cite{amari1998natural}.   

To be more specific, assuming that we have a training set $\CMcal{T}=\{(\bm{x}_{i},y_{i}) \}_{i=1}^{M}$, we start with the estimation of a common set of parameters $\theta^{(0)}$ for the complete training set as
\begin{equation}
\theta^{(0)} = \text{argmin}_{\theta} \sum_{i=1}^{M} \CMcal{L} (\theta, y_i),
\label{eq:theta0}
\end{equation}
where $M$ is the total number of training examples. Next, for each tree $n$, we estimate the individual natural gradients $g_{i}^{(n)}$ with respect to the estimated parameters up to that tree (stage) as
\begin{equation}
g_{i}^{(n)} = \CMcal{I}_{\CMcal{L}} \left( \theta_{i}^{(n-1)} \right) ^{-1} \cdot \nabla_{\theta}\CMcal{L} \left( \theta_{i}^{(n-1)}, y_i \right) \text{,}
\label{eq:g_i_n}
\end{equation}
where, for given $\CMcal{L}$, $\CMcal{I}_{\CMcal{L}} (\theta)$ denotes the Fisher Information that the target variable $y$ carries about the probability distribution $P_{\theta}$ \cite{lehmann2006theory}. For the case of the logarithmic score, the Fisher Information can be written as
\begin{equation}
\CMcal{I}_{\CMcal{L}} (\theta) = \E_{y	\sim P_{\theta}} \left[ \nabla_{\theta}\CMcal{L} (\theta, y) \nabla_{\theta}\CMcal{L} (\theta, y)^{\top} \right] \text{.}
\label{eq:fisher_information}
\end{equation}
Once the natural gradients $g_{i}^{(n)}$ have been estimated, they are used together with the input vectors $\bm{x}_i$ to train the set of base learners of that stage $f^{(n)}$. As a following step, the scaling factor $\rho^{(n)}$ is computed as
\begin{equation}
\rho^{(n)} = \text{argmin}_{\rho} \sum_{i=1}^{M} \CMcal{L} \left( \theta_{i}^{(n-1)} - \rho \cdot f^{(n)}(\bm{x}_{i}), y_{i} \right),
\label{eq:rho_n}
\end{equation}
and the parameters $\theta_{i}^{(n)}$ are updated using
\begin{equation}
\theta_{i}^{(n)} = \theta_{i}^{(n-1)} - \eta \left( \rho^{(n)} f^{(n)}(\bm{x}_{i}) \right),
\label{eq:theta_n}
\end{equation}
where $\eta$ denotes a usual learning rate. Note that when the training is over, we will have two base learners $f_{\mu}^{(n)}$ and $f_{\sigma}^{(n)}$, one estimating the mean $\mu$ and one estimating the standard deviation $\sigma$ (normal distribution), respectively. 

\subsection{Gaussian Process (GP)}
For GP regression \cite{williams2006gaussian}, we assume that we have a set of finite random variables following a joint multivariate Gaussian distribution:
\begin{equation}
f(\bm{x}_{1}),...,f(\bm{x}_{n}) \sim \CMcal{N}(0,\bm{K}),
\label{eq:fxi}
\end{equation}
where $\bm{K}$ expresses the covariance matrix of those random variables. In particular, each element $K_{ij}=C(\bm{x}_{i}, \bm{x}_{j})$ describes the covariance between $f(\bm{x}_{i})$ and $f(\bm{x}_{j})$ and implicitly controls the influence level of the output at $\bm{x}_i$ on the output at $\bm{x}_j$. In this context, the choice of the covariance function $C(\bm{x}_{i}, \bm{x}_{j})$ practically introduces prior knowledge about the expected output. Typical choices for a  covariance function are the radial basis function (RBF), the rational quadratic kernel (RQ), and the periodic kernel, which can be expressed by the following three equations, respectively:
\begin{equation}
C_{\text{RBF}}( \bm{x}_{i}, \bm{x}_{j})= \sigma^{2} \text{exp} \left( - \frac{ \| \bm{x}_{i} - \bm{x}_{j} \| ^2}{2l^{2}} \right),
\label{eq:rbf}
\end{equation}
\begin{equation}
C_{\text{RQ}}( \bm{x}_{i}, \bm{x}_{j})= \sigma^{2} \left(1 + \frac{ \| \bm{x}_{i} - \bm{x}_{j} \| ^2}{2 \alpha l^{2}} \right)^{- \alpha},
\label{eq:rq}
\end{equation}
\begin{equation}
C_{\text{Per}}( \bm{x}_{i}, \bm{x}_{j})= \sigma^{2} \text{exp} \left( - \frac{2}{l^{2}} \sin^{2} \left( \pi \frac{ | \bm{x}_{i} - \bm{x}_{j} | }{p} \right) \right),
\label{eq:periodic_kernel}
\end{equation}
where the parameter $\sigma$ indicates the standard deviation of $f$, the lengthscale $l$ modulates the correlation between neighbor points, $\alpha$ expresses the scale-mixture which regulates the weighting between different lengthscales, and $p$ corresponds to the period of the unknown function. Moreover, combinations of the aforementioned kernels can be also deployed in order to model more complex relationships.

Assuming a Gaussian noise $\epsilon_{i}$ with a variance $\sigma_n$ in our observations, we can mathematically express the output as:
\begin{equation}
y_{i}=f(\bm{x}_{i})+\epsilon_{i}, \quad \epsilon_{i} \sim \CMcal{N}(0,\sigma_{n}^{2}).
\label{eq:y_i}
\end{equation}
Given the noise variance and the parameters of the kernel function are known, it can be proved that the output distribution $y^{*}$ for a new test input $\bm{x}^{*}$ follows also a Gaussian distribution with a mean and a variance that can be estimated by: 
\begin{equation}
\mu (\bm{x}^*)= \bm{k}(\bm{x}^{*})^{\top} \bm{K}^{-1} \bm{y},
\label{eq:mean}
\end{equation}
\begin{equation}
\sigma^{2}(\bm{x}^*) = k(\bm{x}^{*}) - \bm{k}(\bm{x}^{*})^{\top} \bm{K}^{-1} \bm{k}(\bm{x}^{*}).
\label{eq:variance}
\end{equation}
Here, assuming that $\bm{\Sigma}$ denotes the covariance matrix of the noise-free system and $\bm{I}$ is the $M \times M$ identity matrix, their combination $\bm{K}=\bm{\Sigma} + \sigma_n \bm{I}$ expresses the covariance matrix of the full system. The vector $\bm{y} = [y_1, ..., y_M]^{\top}$ contains the observed target variables, $\bm{k}(\bm{x}^{*}) = [C(\bm{x}_1, \bm{x}^*),..., C(\bm{x}_M, \bm{x}^*)]^\top$ expresses the covariances between the input vectors $\bm{x}_i$ and the test input $\bm{x}^{*}$, and $k(\bm{x}^{*})$ indicates the prior variance of $\bm{x}^{*}$. 


The training of the model, i.e., the estimation of model parameters, is done by the minimization of the negative log marginal likelihood of  $p(\bm{y}, \textbf{X})$, where $\textbf{X} = [\bm{x}_1, ..., \bm{x}_M]^{\top}$. The minimization is solved using the Adam optimizer \cite{kingma2014adam}.

\subsection{Lower Upper Bound Estimation (LUBE)}
LUBE is a neural network based algorithm, which, as the name implies, estimates the lower and upper bound of the PIs associated with a point forecast \cite{khosravi2010lower}. Most of the times, the neural network comprises one hidden layer and two output neurons corresponding to the lower and upper bounds of the PIs, $L(\bm{x}_i)$ and $U(\bm{x}_i)$, respectively. However, contrary to the traditional neural networks typically using the mean square error as a cost function, LUBE considers a combination of two terms, namely the number of data points lying within the yielded PIs and the total width of those PIs. Those two terms can be expressed as: 
\begin{equation}
\text{PICP} = \frac{1}{M} \sum_{i=1}^{M} c_{i},
\label{eq:PIPC}
\end{equation}
\begin{equation}
\text{PINAW} = \frac{\frac{1}{M} \sum_{i=1}^{M} \left( U\left( \bm{x}_i \right) - L\left( \bm{x}_i \right) \right)}{R}.
\label{eq:PINAW}
\end{equation}

The PI coverage probability (PICP) is essentially the percentage of the output points covered by the generated PIs. In this regard, $c_{i}=1$ if $y_i \in [L(\bm{x}_i), U(\bm{x}_i)]$ and $c_{i}=0$ otherwise. In addition, the PI normalized average width (PINAW) is used to balance the influence of PICP so that the PIs do not become very large. The normalization variable $R$ corresponds to the maximum value of the target variable. Now, the cost function of the neural network, also called coverage width-based criterion (CWC), can be expressed as: 
\begin{equation}
\text{CWC} = \text{PINAW} \left( 1 + \gamma \left( \text{PICP} \right) e^{- \eta (\text{PICP} - \mu)} \right),
\label{eq:CWC}
\end{equation} 
where the hyperparameter $\mu$ denotes the confidence level of the PIs while $\eta$ is as a penalty factor magnifying small differences between PICP and $\mu$. The term $\gamma (\text{PICP})$ is 1 while training the network and is predominantly employed for assessing the PIs after the training. In that case, $\gamma = 1$ if $\text{PICP} \geq \mu$ and $\gamma = 0$ otherwise. Finally, the simulated annealing algorithm is used for the estimation of the weights and the biases of the neural network, since the CWC is nonlinear and discontinuous \cite{aarts1989simulated}.

It is worth mentioning that LUBE yields only PIs and thus, its performance will be assessed using only the probabilistic metrics described in the next section.


\subsection{Validation metrics}
In this section, we present the main metrics concerning the accuracy, bias, reliability, and sharpness of a probabilistic forecast.

The accuracy and the bias refer to the point forecasts and express how much they deviate from the real power values. The most commonly used metrics are the mean absolute error (MAE), the root mean square error (RMSE), and the mean bias error (MBE). Those metrics can be calculated as:
\begin{equation}
\text{MAE} = \frac{1}{M} \sum_{i=1}^{M} \lvert y_{i}^{*} - y_i \rvert ,
\label{eq:mae}
\end{equation} 
\begin{equation}
\text{RMSE} = \sqrt{\frac{1}{M} \sum_{i=1}^{M} \left( y_{i}^{*} - y_i \right) ^{2}} , 
\label{eq:rmse}
\end{equation} 
\begin{equation}
\text{MBE} = \frac{1}{M} \sum_{i=1}^{M} \left( y_{i}^{*} - y_i \right).
\label{eq:mbe}
\end{equation} 

Two important properties of a probabilistic forecast are its reliability and its sharpness. The former expresses whether the real power values could have been drawn from the predicted distribution whereas the latter expresses how concentrated the predicted distribution is \cite{pinson2007non}. A typical metric to evaluate reliability is the histogram of the probability integral transform (PIT). The PIT value of an observation $y_i$ is estimated by applying the predicted cumulative distribution function (CDF) $\hat{F}$ to that observation as
\begin{equation}
\text{PIT}_i = \hat{F}(y_i).
\label{eq:pit}
\end{equation} 
By applying the predicted CDF to all observations, we acquire the histogram of PIT, which should be as close as possible to a uniform histogram \cite{gneiting2014probabilistic}. U-shaped histograms imply overconfidence whereas concave-shaped histograms imply the opposite. Another metric measuring reliability is PICP which can be calculated using \eqref{eq:PIPC}. Similarly, a commonly used metric for sharpness is PINAW, as described in \eqref{eq:PINAW}.  

Due to the variety of forecasting metrics addressing different model properties, it is difficult to generally benchmark different models by jointly comparing the various metrics. Therefore, the negatively oriented continuous ranked probability score (CRPS) is employed, which takes into consideration the aforementioned desired properties of a model and generates a score expressed in the units of the forecast quantity. It is defined as
\begin{equation}
\text{CRPS}_i = \int_{-\infty}^{\infty} \left( \hat{F}(y) - \mathds{1} \left( y - y_i \right) \right)^2 dy \text{,}
\label{eq:crps}
\end{equation}
where $\mathds{1}$ represents the Heaviside function, which is 1 if its argument is nonnegative and 0 otherwise.  

\section{Model interpretability}
Probabilistic forecasting models are predominantly complex black box models that lack transparency. In the second stage of the proposed framework, we aim at understanding the model and its learned feature relationships. In particular, we focus on interpreting the NGBoost model using the SHAP values. Note that the term interpretability refers to the ability of the model to communicate why and how a prediction was made using feature attribution values, which basically describe the contribution of each feature to the final output.

\subsection{SHAP values}
As mentioned, the interpretability of a model is usually defined by a set of feature attribution values that quantify the influence of each input feature on the model output. Those values can describe the influence of a feature as a whole (global) or they can correspond to a single prediction (local). For tree based models, the gain \cite{breiman1984classification}, the split count \cite{chen2016xgboost}, the permutation \cite{ishwaran2007variable}, and the SHAP \cite{lundberg2017unified} approach are currently the main approaches for estimating the global feature importance. In the case of local explanations, the SHAP and the Saabas method \cite{saabas2014interpreting} form the two unique alternatives to date. However, the gain, split count, and Saabas approaches are inconsistent, meaning that the estimated feature attributions do not always align with the actual feature impact on a prediction \cite{lundberg2020local}. Furthermore, although permutation approaches are consistent, they only provide global feature importance. On the contrary, the SHAP approach is able to deliver local explanations while providing theoretical guarantees about the method's consistency based on game theory \cite{lundberg2020local}. For those reasons, we opt for the SHAP method in order to interpret the predictions generated in the first stage of the proposed framework.

In order to estimate the individualized feature attributions, it is assumed that $g$ is an explanation model, which can be seen as an interpretable approximation of our initial model $f$. $g$ can be written as a linear combination of the feature attributions $\phi_d \in \mathbb{R}$ with the binary coefficients $z_{d}^{\prime} \in \{ 0,1 \}$. Here, $z_{d}^{\prime} = 1$ when the feature $d$ is observed whereas $z_{d}^{\prime} = 0$ when the feature is unknown, i.e., has not been observed. In this regards, $g$ can be expressed as:
\begin{equation}
g(\bm{z}^{\prime}) = \phi_{0} + \sum_{d=1}^{D} \phi_{d} \cdot z_{d}^{\prime},
\label{eq:gz}
\end{equation}
where $\phi_{0}$ denotes the initial expectation about the model output without observing any feature (base value) and $D$ indicates the total number of features.

The SHAP method estimates the influence of a feature by observing how the model behaves with and without that feature. To do so, it casts the problem of feature attribution to a cooperative game theory problem, where each player (feature) contributes differently to the game. As proved in \cite{lundberg2018explainable}, SHAP values are the unique consistent and local accurate solution subject to the missingness property (the attribution of missing features is zero) for calculating the contribution of each player to the final game result. 

In particular, a function $h_{\bm{x}}$ is initially defined in order to map the binary vector $\bm{z}^{\prime}$ into the original input space. Given the function $h_{\bm{x}}$, the influence of observing a feature can be estimated by assessing the value of $f(h_{\bm{x}}(\bm{z}^{\prime}))$, which is the expected value of the model $f$ conditioned on a subset of features $\CMcal{S} \subseteq \CMcal{Z}$ ($\CMcal{Z}$ being the set containing all input features). This can be written by defining $f_{\bm{x}}(\CMcal{S}) = f(h_{\bm{x}}(\bm{z}^{\prime})) = \E[f(\bm{x}) | \bm{x}_{\CMcal{S}}]$, where $\bm{x}_{\CMcal{S}}$ indicates the values of the features defined by $\CMcal{S}$. In order to obtain the correct contribution of a feature, we need to consider the interaction effects with the other features and thus, its influence is estimated over all possible subsets of features. Those different combinations result in a computationally challenging calculation:
\begin{equation}
\phi_{d} = \sum_{\CMcal{S} \subseteq \CMcal{Z} \setminus \{ d \}} \frac{|\CMcal{S}|! (|\CMcal{Z}| - |\CMcal{S}| -1)! }{|\CMcal{Z}|!} \left[ f_{\bm{x}}(\CMcal{S} \cup \{d\} ) - f_{\bm{x}}(\CMcal{S}) \right].
\label{eq:phi_d}
\end{equation}

While SHAP values can be approximated for black box models, e.g., neural networks, support vector machines \cite{lundberg2017unified}, tree based models enjoy the exact calculation of SHAP values in polynomial instead of exponential time based on the work in \cite{lundberg2020local}. Under those conditions, we can further support our motivation to combine a tree based model, such as NGBoost, with the SHAP method in order to introduce a computationally efficient interpretable probabilistic forecasting framework.

\subsection{SHAP interaction values}
The feature attributions calculated by \eqref{eq:phi_d} determine the influence of individual features without revealing any information about the various interactions with other features. To capture pairwise interactions between features, the SHAP interaction values have been introduced in \cite{lundberg2020local}. To this end, the classic Shapley values have been extended to a Shapley interaction index, which can be calculated by
\begin{equation}
\Phi_{d_1, d_2} = \sum_{\CMcal{S} \subseteq \CMcal{Z} \setminus \{ d_1, d_2 \}} \frac{|\CMcal{S}|! (|\CMcal{Z}| - |\CMcal{S}| -2)! }{2 (|\CMcal{Z}|-1)!} \nabla_{d_{1} d_{2}} (\CMcal{S}),
\label{eq:phi_d1_d2}
\end{equation}
where
\begin{equation}
\begin{array}{ll}
\nabla_{d_{1} d_{2}}(\CMcal{S}) = f_{\bm{x}}(\CMcal{S} \cup \{d_1, d_2\} ) + f_{\bm{x}}(\CMcal{S}) \\ \quad \quad \quad \quad - f_{\bm{x}}(\CMcal{S} \cup \{d_1 \} ) - f_{\bm{x}}(\CMcal{S} \cup \{d_2\} ) .
\end{array}
\label{eq:grad_d1_d2}
\end{equation}
Based on \eqref{eq:phi_d1_d2}, the SHAP interaction value between $d_1$ and $d_2$ is divided into two equal parts, $\Phi_{d_1,d_2}$ and $\Phi_{d_2,d_1}$. Hence, the total interaction value between the features $d_1$ and $d_2$ is calculated by adding those two terms, i.e., $\Phi_{d_1,d_2} + \Phi_{d_2,d_1}$. 

Consequently, the SHAP interaction value can be inferred as the difference between the SHAP values of feature $d_1$ when the feature $d_2$ is observed and not observed. In this context, the SHAP interaction values can reveal interesting pairwise relationships that otherwise would be missed, e.g., physics based relationships. 

Similarly to the SHAP values, the SHAP interaction values can be calculated in polynomial time following the work of \cite{lundberg2020local}.

\section{Results}

\subsection{Comparative results}

In this and the following subsections, we focus on the first stage of the proposed forecasting framework and in particular, we thoroughly examine the performance of NGBoost in probabilistic PV forecasting while taking into account the influence of the seasons. To do so, we compare the proposed NGBoost with LUBE (only for PIs), GP, and persistence (only for point forecasts). For the persistence model,  which is included as a benchmark model with minimal model complexity, we assume that the predicted power at a certain point in time will be the same as the power generated the previous day at the exact same time. As for the machine learning approaches, different models of the same algorithm are developed based on different sets of hyperparameters. Specifically, we vary the depth of the tree, the learning rate, the boosting iterations, the distribution, and the score for NGBoost, the number of neurons and the penalty factor $\eta$ for LUBE (increments of 10), and the kernel function for GP. The exact values of those hyperparameters can be found in Table~\ref{tab:hyperparameters}. As a result, numerous models corresponding to all possible combinations of hyperparameteres were developed. Those models were thoroughly evaluated using both point forecasting (MAE, RMSE, MBE) as well as probabilistic metrics (PICP, PINAW, CRPS).

\begin{table}[b!]
\centering
\caption{Hyperparameters of the different models.}
\label{tab:hyperparameters}
\begin{tabular}{ll}
\hline
NGBoost & depth=\{3,4,5\}                             \\
        & learning rate = \{0.01, 0.05, 0.1\}         \\
        & boosting iterations = \{100, 500, 1000\}    \\
        & distribution = \{Gaussian, Laplace\}        \\
        & score = \{log, CRPS\}                       \\ \hline
LUBE    & neurons = \{10, 20, ..., 100\}              \\
        & $\eta$ = \{10, 20, ..., 90\}                \\ \hline
GP      & kernel = \{RBF. RQ, Per, RQ + Per, RQ $\cdot$ Per\} \\ \hline
\end{tabular}
\end{table}

Based on those models, the following observations have been made. The depth of the tree seems to have no remarkable influence on the NGBoost performance while 500 boosting iterations combined with a learning rate of 0.01 showed the best performance overall. Regarding the scoring rule, the logarithmic and CRPS score perform almost equally well in all the seasons apart from winter, where CRPS demonstrated a significant decay in accuracy ($\approx$50\%). Lastly, the Gaussian distribution generates slightly more accurate results than Laplace (around 10\% on average). Furthermore, LUBE models were developed for all the possible combinations of the number of neurons and the parameter $\eta$ shown in Table~\ref{tab:hyperparameters}. In this context, the LUBE models with 20 neurons and a penalty factor of 50 show the best performance. Concerning GP, the RQ kernel yields marginally better results than the rest of the kernels. 

\begin{figure}[t!] 
\includegraphics[width=\linewidth]{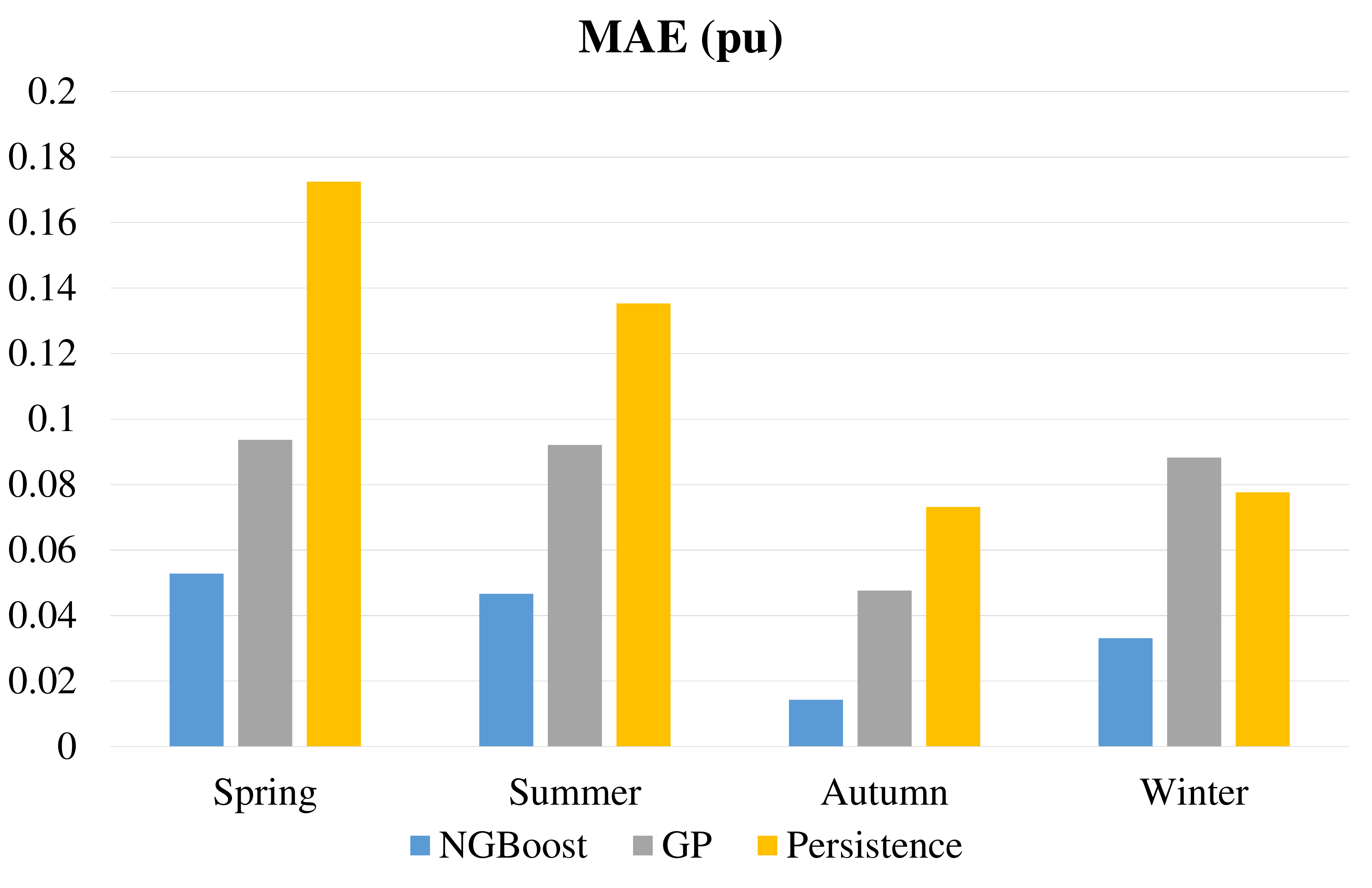}
\caption{Average MAE of the different test sets throughout the year.}
\label{fig:mae}
\end{figure}

\begin{figure}[t!] 
\includegraphics[width=\linewidth]{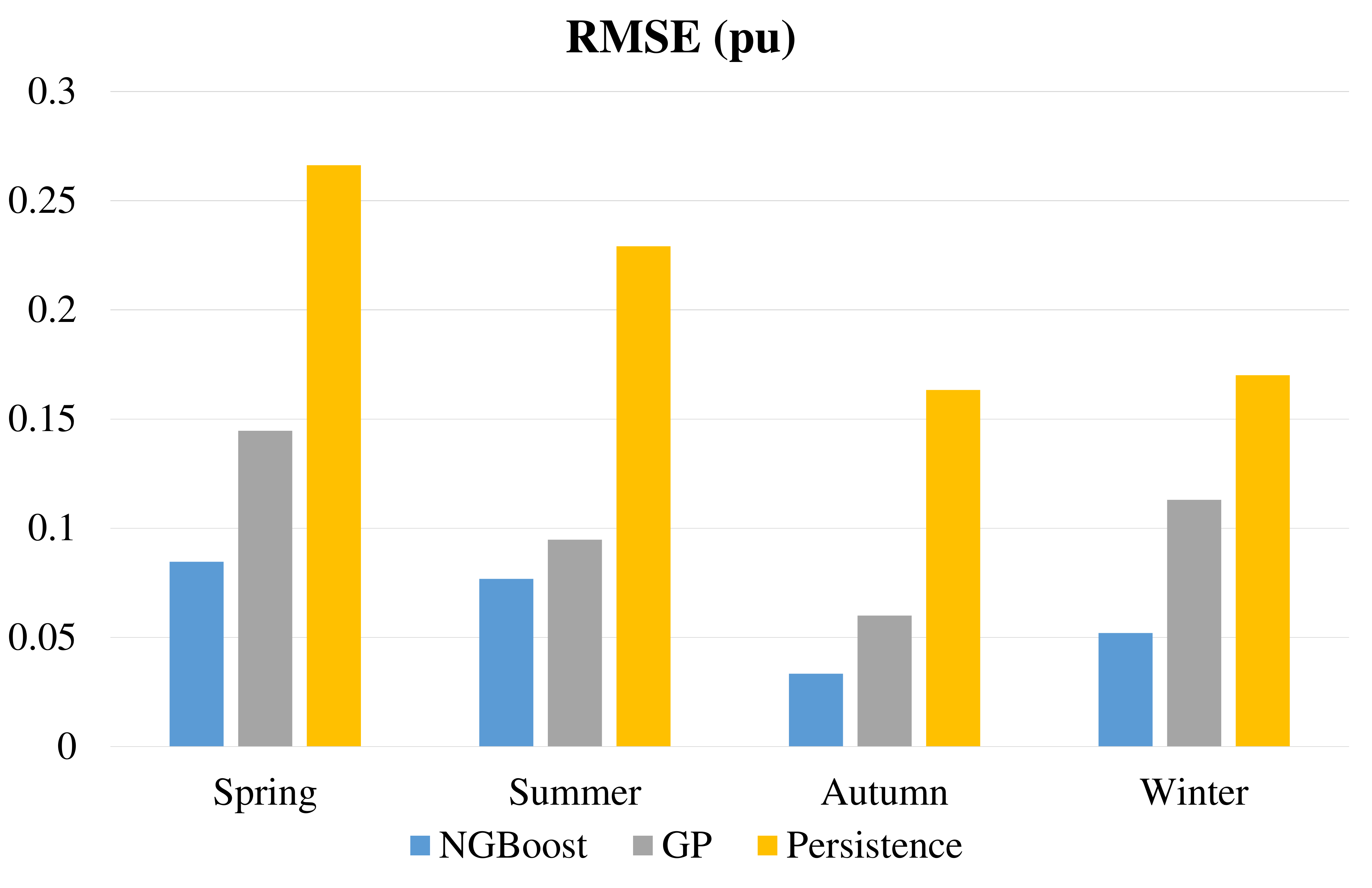}
\caption{Average RMSE of the different test sets throughout the year.}
\label{fig:rmse}
\end{figure}

\begin{figure}[t!] 
\includegraphics[width=\linewidth]{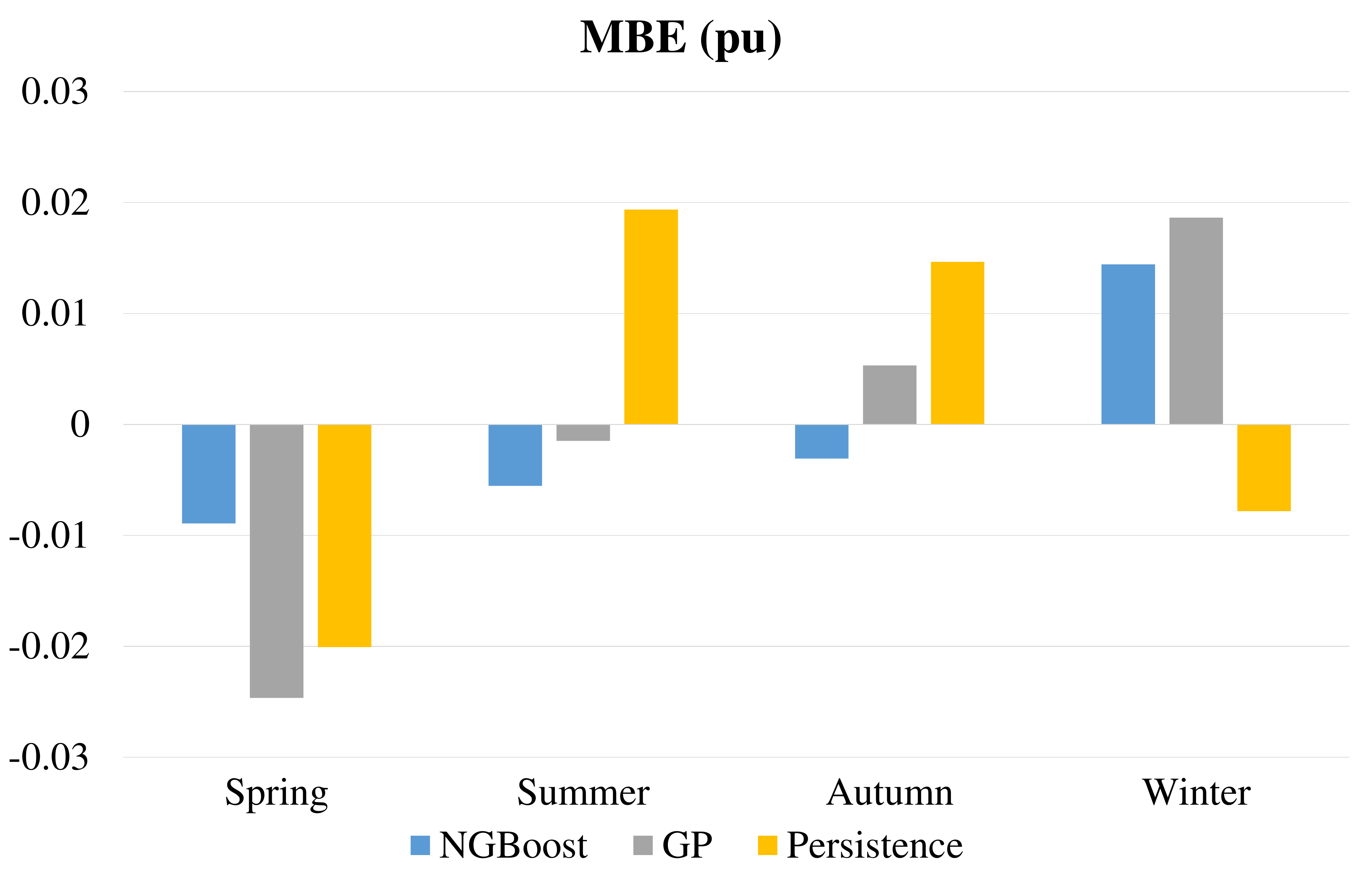}
\caption{Average MBE of the different test sets throughout the year.}
\label{fig:mbe}
\end{figure}

\begin{figure}[t!] 
\includegraphics[width=\linewidth]{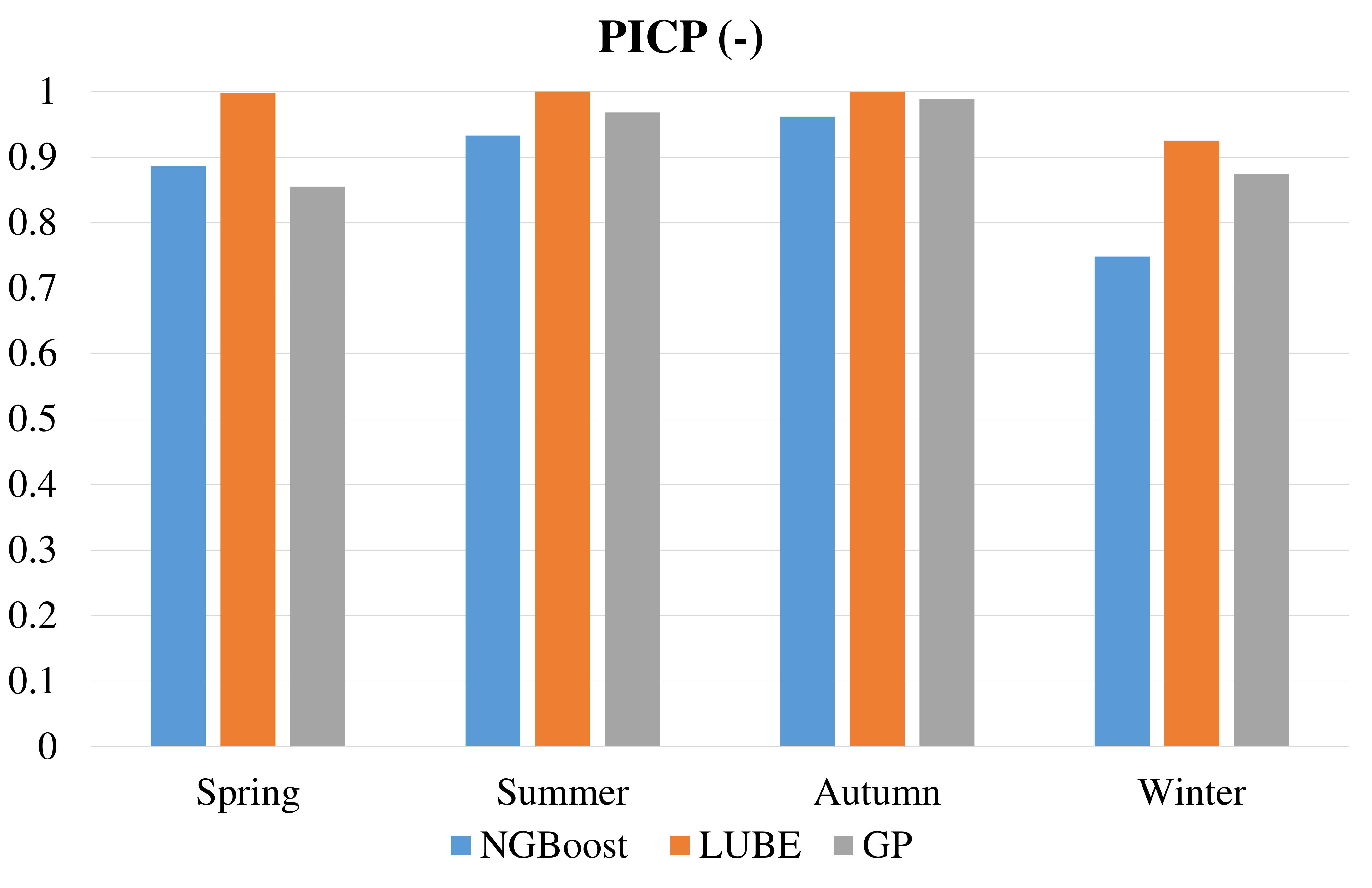}
\caption{Average PICP of the different test sets throughout the year.}
\label{fig:picp}
\end{figure}

\begin{figure}[t!] 
\includegraphics[width=\linewidth]{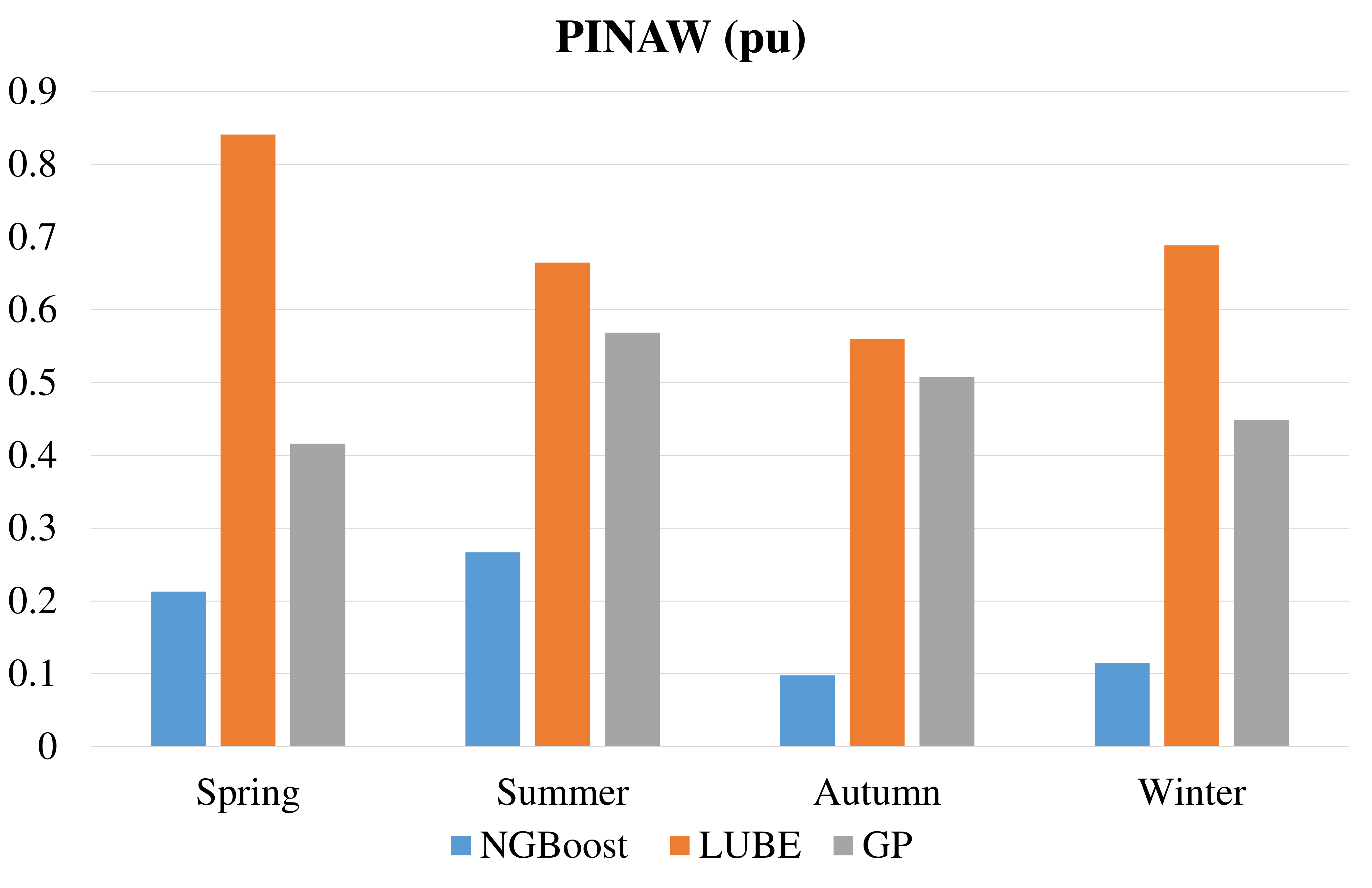}
\caption{Average PINAW of the different test sets throughout the year.}
\label{fig:pinaw}
\end{figure}

\begin{figure}[t!] 
\includegraphics[width=\linewidth]{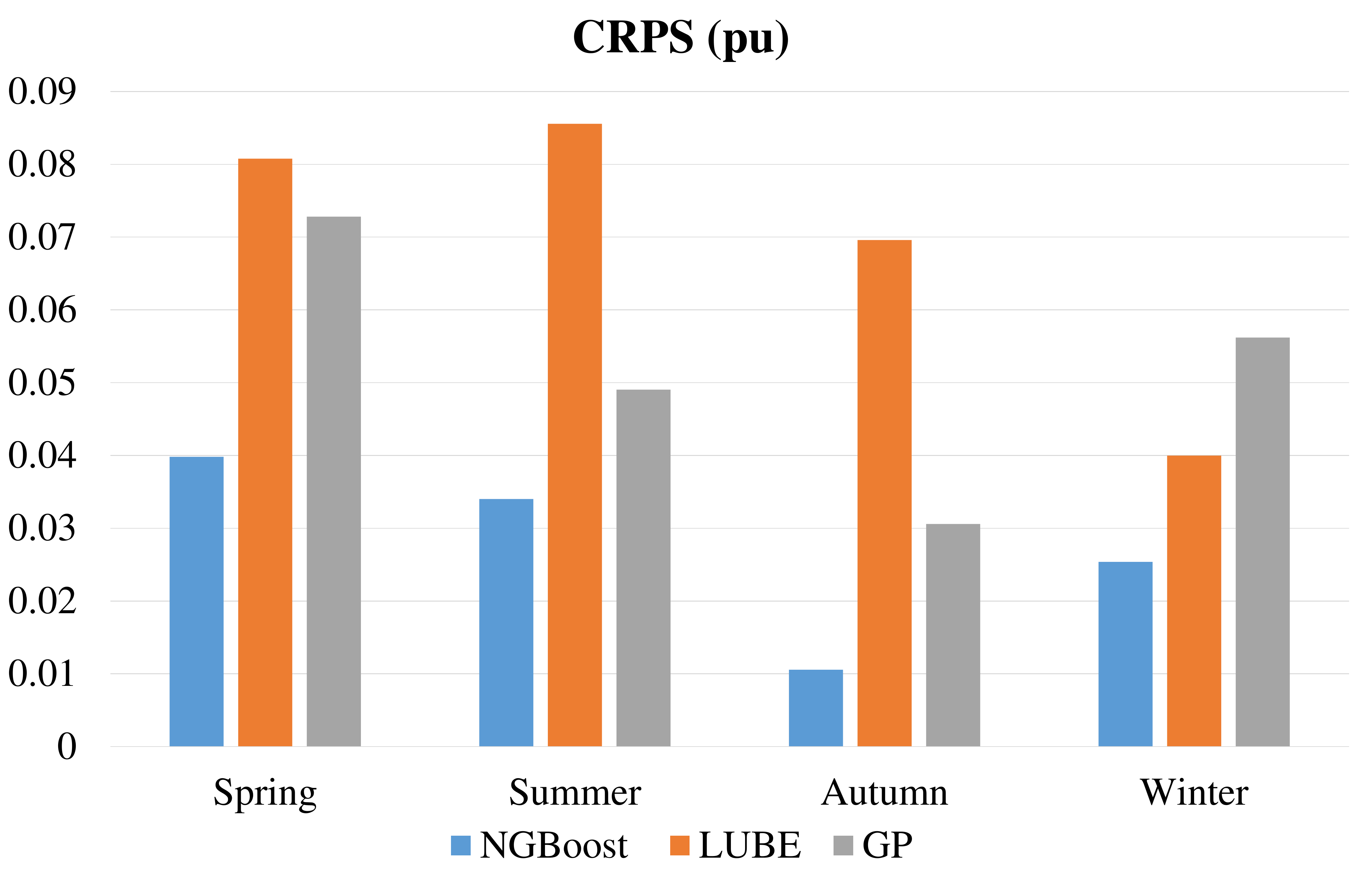}
\caption{Average CRPS of the different test sets throughout the year.}
\label{fig:crps}
\end{figure}

Due to the large number of models, the one with the best overall performance is selected for each method and is presented in the comparative plots of Fig.~\ref{fig:mae}-\ref{fig:crps}. Those results refer to the average values of PVP1 and PVP2, which were scaled by the nominal power. Moreover, the hyperparameters of the presented models are:
\begin{itemize}
\item \textit{NGBoost:} \{depth=3, learning rate=0.01, boosting iterations=500, score=log, distribution=Gaussian\}
\item \textit{LUBE:} \{neurons=20, $\eta$=50\}
\item \textit{GP:} \{kernel=RQ\}
\end{itemize}


Starting with the metrics related to point forecasts (MAE, RMSE, MBE), NGBoost and GP generate significantly better results than the persistence model for all the four seasons. Only for winter days, GP shows a small decay in MAE and MBE. Moreover, NGBoost outperforms GP with respect to the point forecasts, regardless of the time of the year, as clearly observed in Fig.~\ref{fig:mae}-\ref{fig:mbe}. Specifically, NGBoost leads to an overall decrease of around 56\% and 40\% in MAE and RMSE, respectively, compared to GP, while the same metrics reached an overall 70\% drop compared to persistence. At the same time, NGBoost shows almost zero bias for all the seasons whereas GP seems to underestimate the PV generation during spring and overestimate it during winter. Thus, NGBoost can be considered as an adequate choice for accurate point predictions.


Regarding the PIs, LUBE exhibits the highest PICP (Fig.~\ref{fig:picp}) even though its hyperparameter $\mu$ was set to 95\% (confidence level of the included PIs). This comes at a cost of a higher PINAW (Fig.~\ref{fig:pinaw}), which is significantly higher than the ones of NGBoost and GP. Furthermore, the proposed algorithm yields on average 64\% and 75\% narrower PIs throughout the year than GP and LUBE, respectively. As for CRPS (Fig.~\ref{fig:crps}), it can be clearly observed that the overall performance of NGBoost in probabilistic forecasting is remarkably better than that of GP and LUBE reaching an average drop of around 50\%. Importantly, NGBoost seems quite robust with respect to the various seasons. In contrast, the performance of GP and LUBE is influenced by the time of the year as revealed in Fig.~\ref{fig:mae}-\ref{fig:crps}.

%



\textit{Training time:} In Table \ref{tab:training_time}, we present the average training times per algorithm. It is worth emphasizing that the NGBoost models can be trained within less than 2 minutes whereas the other two algorithms require around 25 minutes each. The fast training time is of significant importance for the development of a machine learning model in practice, as initially, different features and hyperparameters should be tried out in order to find the optimal combination of them. All models were developed in Python and the training was performed by a personal computer with an Intel Core i5-8500 CPU, 3.00 GHz processor and 8 GB of RAM. The code can be found in the GitHub repository of \cite{mitre_github}.

\begin{table}[t!]
\centering
\caption{Average training time in seconds. }
\label{tab:training_time}
\begin{tabular}{lccc}
\hline
\multicolumn{1}{c}{} & \multicolumn{1}{c}{\textbf{NGBoost}} & \textbf{LUBE} & \textbf{GP} \\ \hline
Training time (s)    & 109                                    & 1439           & 1791     \\ \hline
\end{tabular}
\end{table}

\subsection{Probabilistic forecasting}
\begin{figure}[b!] 
    \subfloat[Spring day.\label{fig:pv_spring}]{%
       \includegraphics[width=0.5\linewidth]{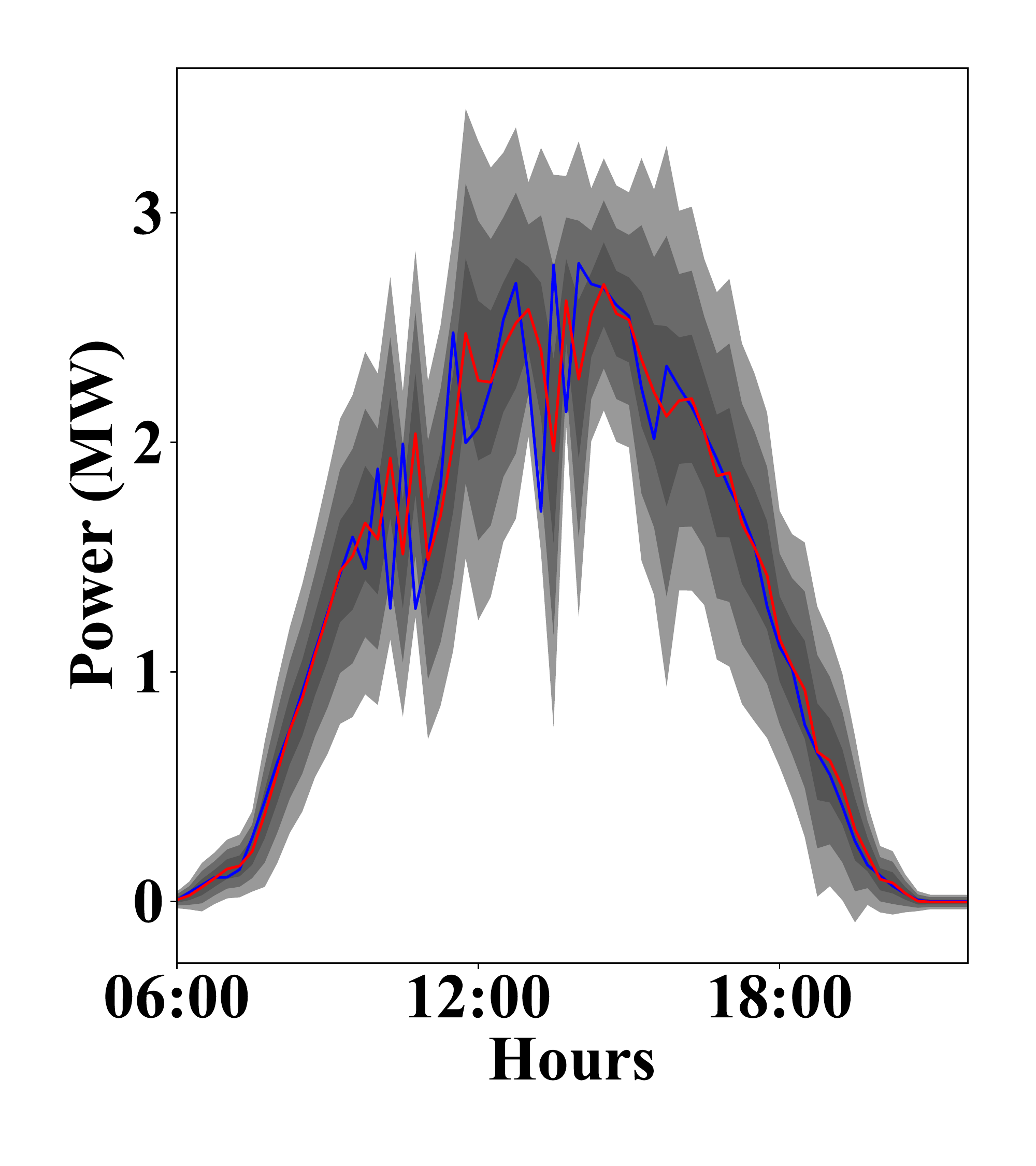}}
	\subfloat[Summer day.\label{fig:pv_summer}]{%
       \includegraphics[width=0.5\linewidth]{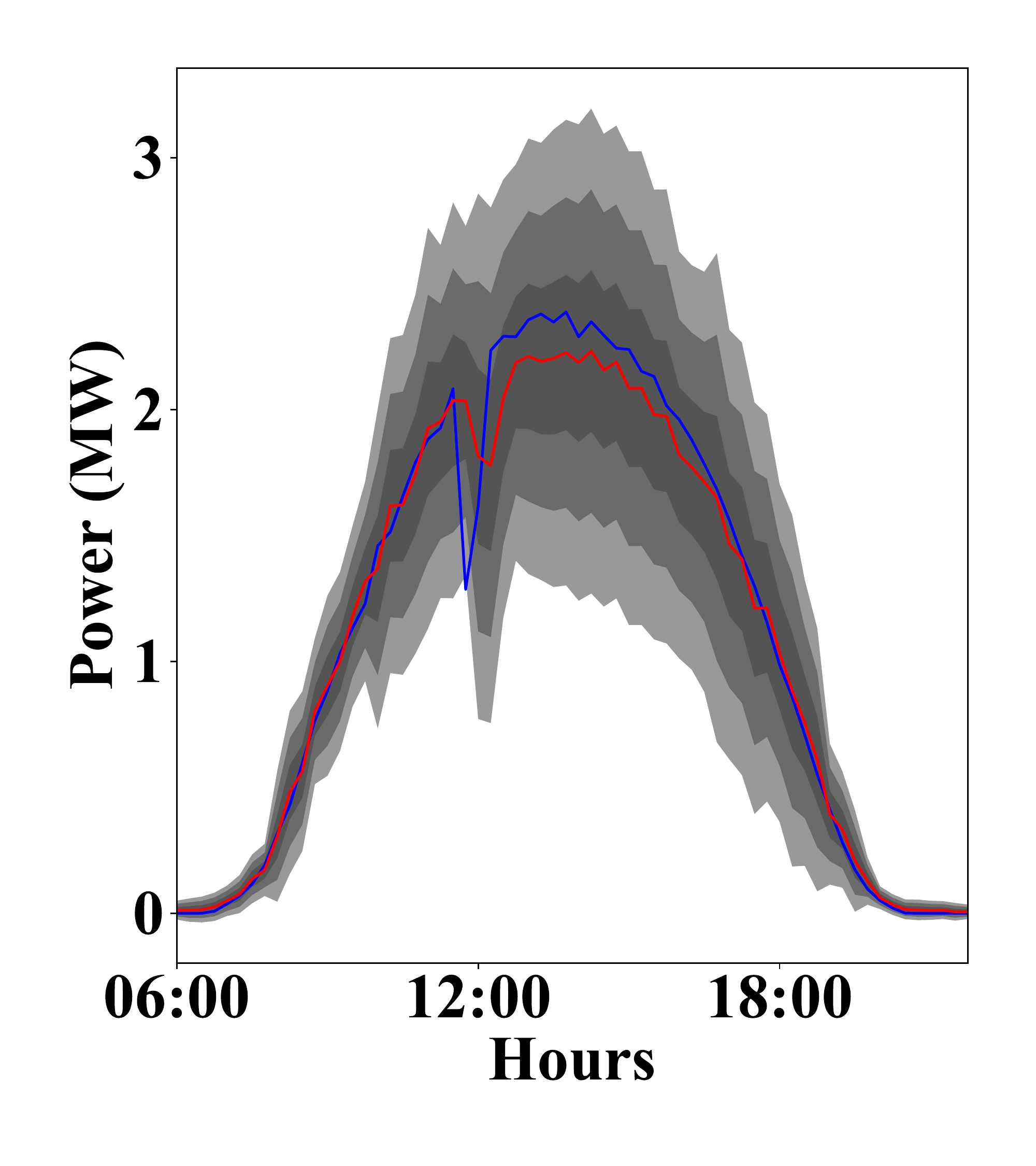}}
       \
    \subfloat[Autumn day.\label{fig:pv_autumn}]{%
       \includegraphics[width=0.5\linewidth]{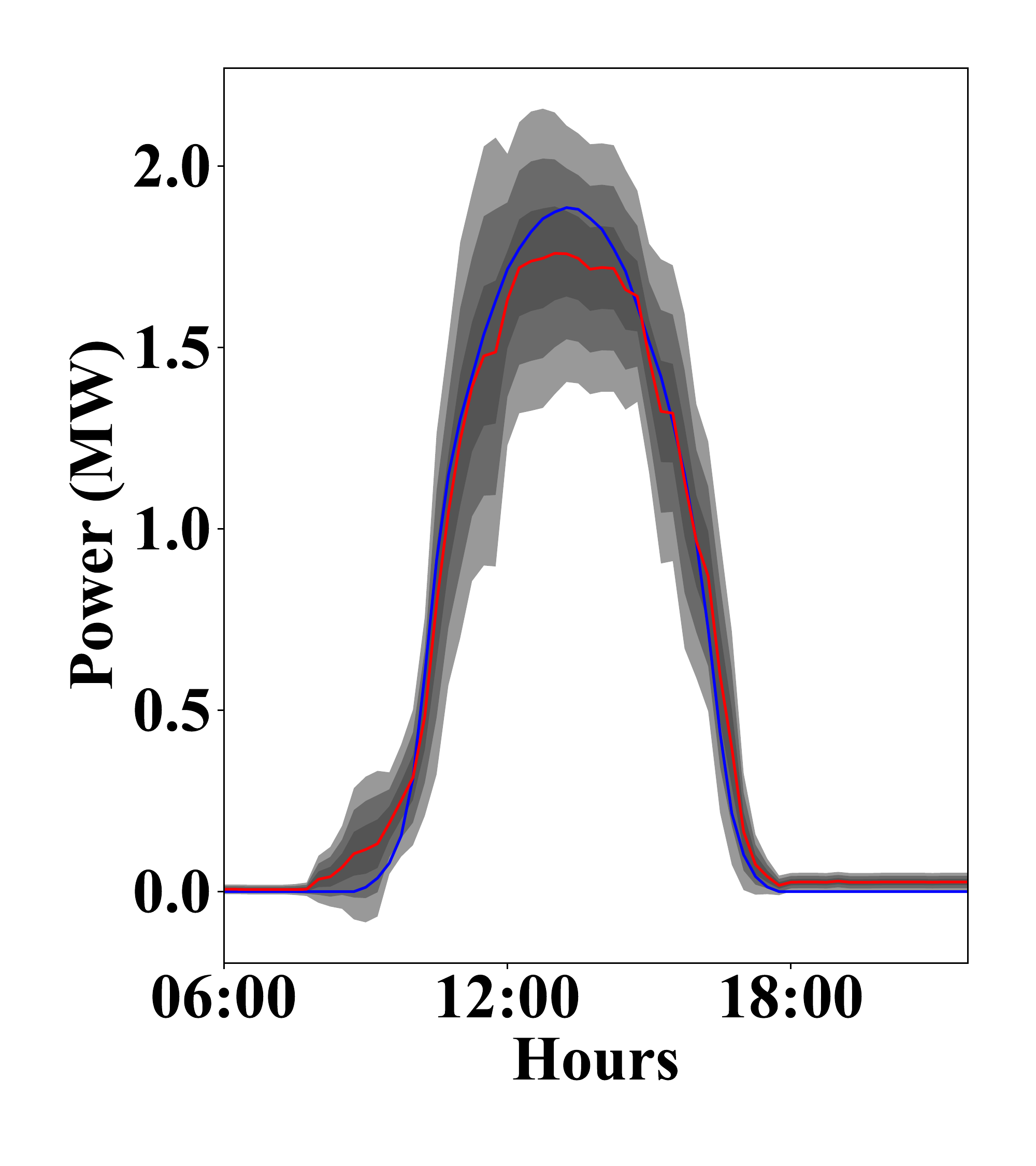}}
	\subfloat[Winter day.\label{fig:pv_winter}]{%
       \includegraphics[width=0.5\linewidth]{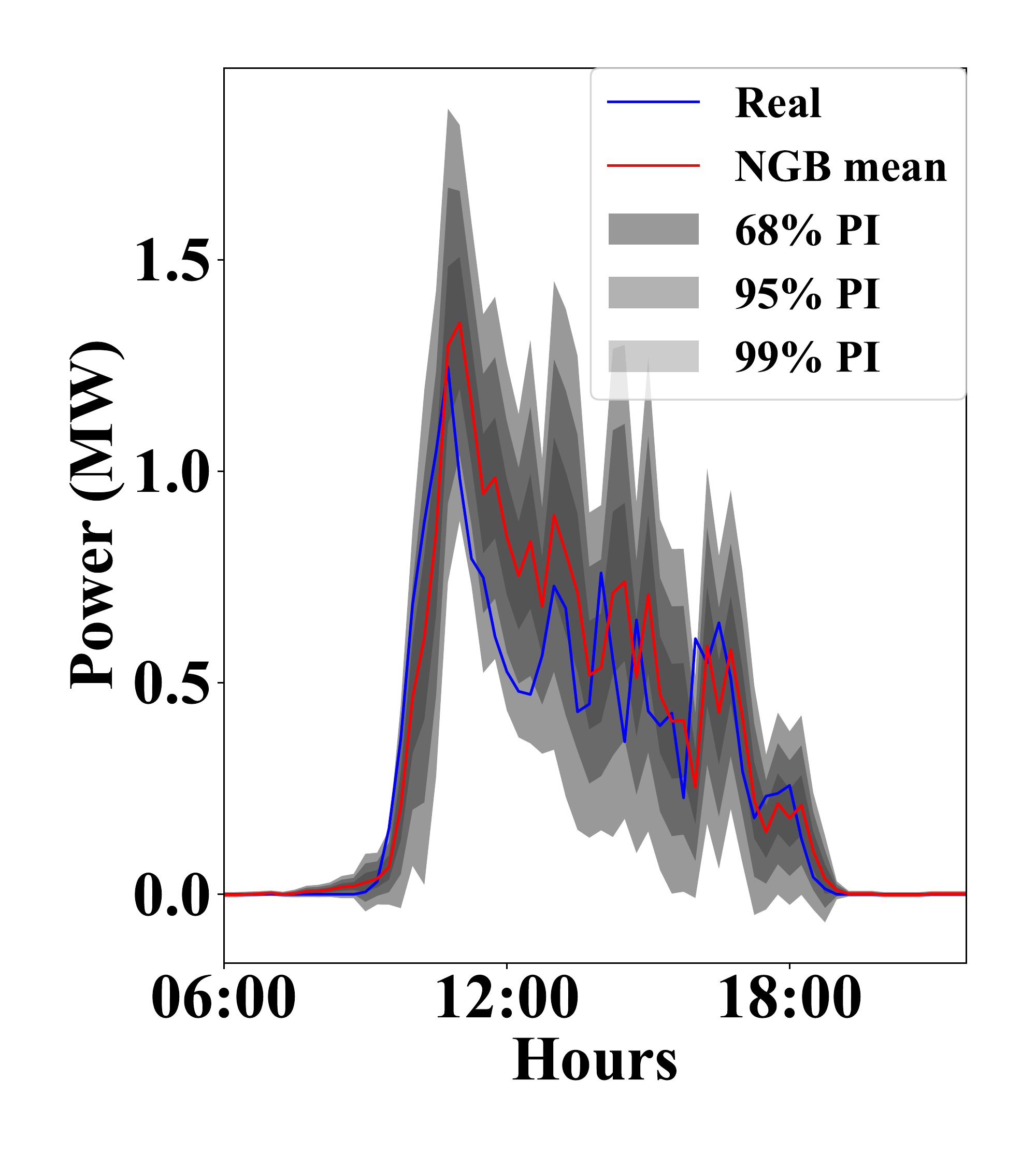}}

\caption{PVP1: Day-ahead PV power forecasts using NGBoost. The individual plots illustrate the model output (red curve) alongside with the real power values (blue curve) as well as the corresponding PIs (gray shading regions).}
\label{fig:pv_examples}
\end{figure}

Fig.~\ref{fig:pv_examples} shows four indicative results as generated by the NGBoost algorithm using data from PVP1. In particular, those four plots refer to the day-ahead forecasts, which were randomly selected by the corresponding test sets. As depicted in Fig.~\ref{fig:pv_examples}, each plot illustrates a completely different generation pattern due to seasonal variations. Yet the NGBoost algorithm was able to yield highly accurate and sharp probabilistic forecasts, where the point forecasts lie always within the boundaries of the predicted distributions. Since we opted for a Gaussian distribution, we plot approximately the 68\%, 95\%, and the 99\% PIs, which correspond to the values less than one, two, and three standard deviations away from mean, respectively.

\begin{figure}[t!] 
\includegraphics[width=\linewidth]{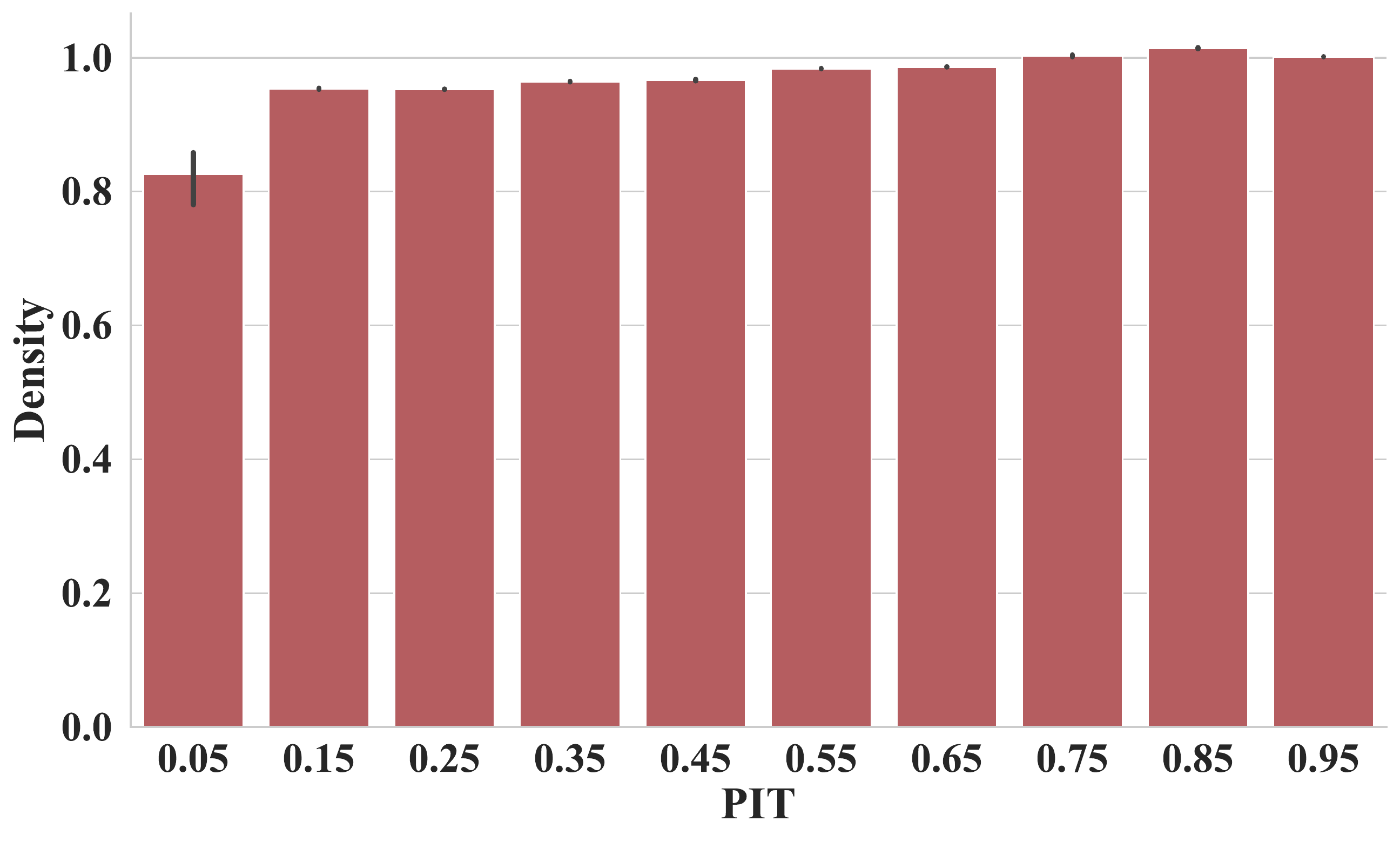}
\caption{Combined PIT histogram of the NGBoost models.}
\label{fig:pit}
\end{figure}

In order to assess the reliability of the two NGBoost models corresponding to PVP1 and PVP2, respectively, the common histogram of their PITs as defined in \eqref{eq:pit} is presented in Fig.~\ref{fig:pit}. Note that the power values were scaled using the respective nominal power. For PIT values above 0.5, the models exhibit almost perfectly reliable forecasts, whereas there is a small tendency of underestimating the density of the bins in the lower half. Nevertheless, the yielded values are quite close to a perfectly reliable model.

To summarize, the proposed model is able to yield reliable and more accurate predictions compared to two of the state-of-the-art approaches. In this regard, PV park operators, system operators, and power traders could benefit from the high forecasting accuracy by optimally placing their bids in the short term energy market. As a result, the system operators would require less balancing energy to be activated and the PV park operators or traders would increase their profits by minimizing the imbalance costs. Moreover, the proposed method offers a forecasting solution that minimizes the expert knowledge required for its implementation. On the contrary, state-of-the-art approaches are often characterized by very complex models with long training times while the source code is mostly not available. Therefore, their implementation in practice may be challenging. To this end, we have published our code as open source so that anyone can test it and apply it to the corresponding problem without many modifications needed.

\subsection{Interpretation of point forecasts}
\begin{figure}[t!] 
\includegraphics[width=\linewidth]{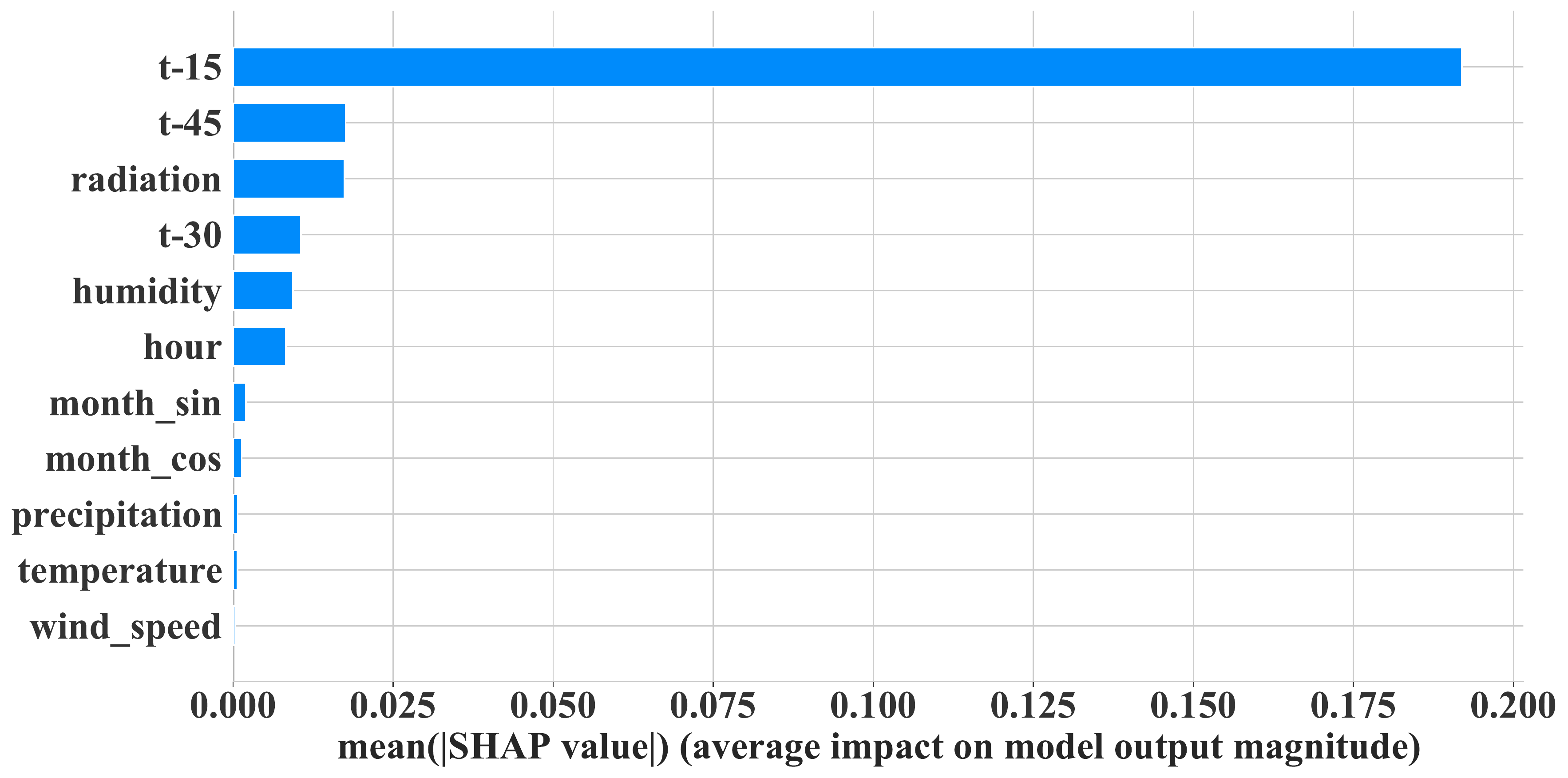}
\caption{PVP1: Global feature importance of the point forecast model. Each bar plot correspond to the global importance of an individual feature.}
\label{fig:bar_yearly_mean_ing}
\end{figure}

In the following sections, we rigorously study the results of the second stage of the proposed forecasting framework which aims at interpreting and understanding the output of the probabilistic forecasting model. First, we focus on the point forecasts and especially, how the derived model exploits the different features to make a prediction. As mentioned, the SHAP method provides both global and local explanations. Starting with the global ones, the average contribution of each feature is demonstrated in Fig.~\ref{fig:bar_yearly_mean_ing}. It is clearly visible that the past power values and the radiation dominate the model, while humidity and time of the day have smaller contributions. The month and the weather information about precipitation, temperature, and wind speed have a negligible effect on the model output. 

\begin{figure}[t!] 
\includegraphics[width=\linewidth]{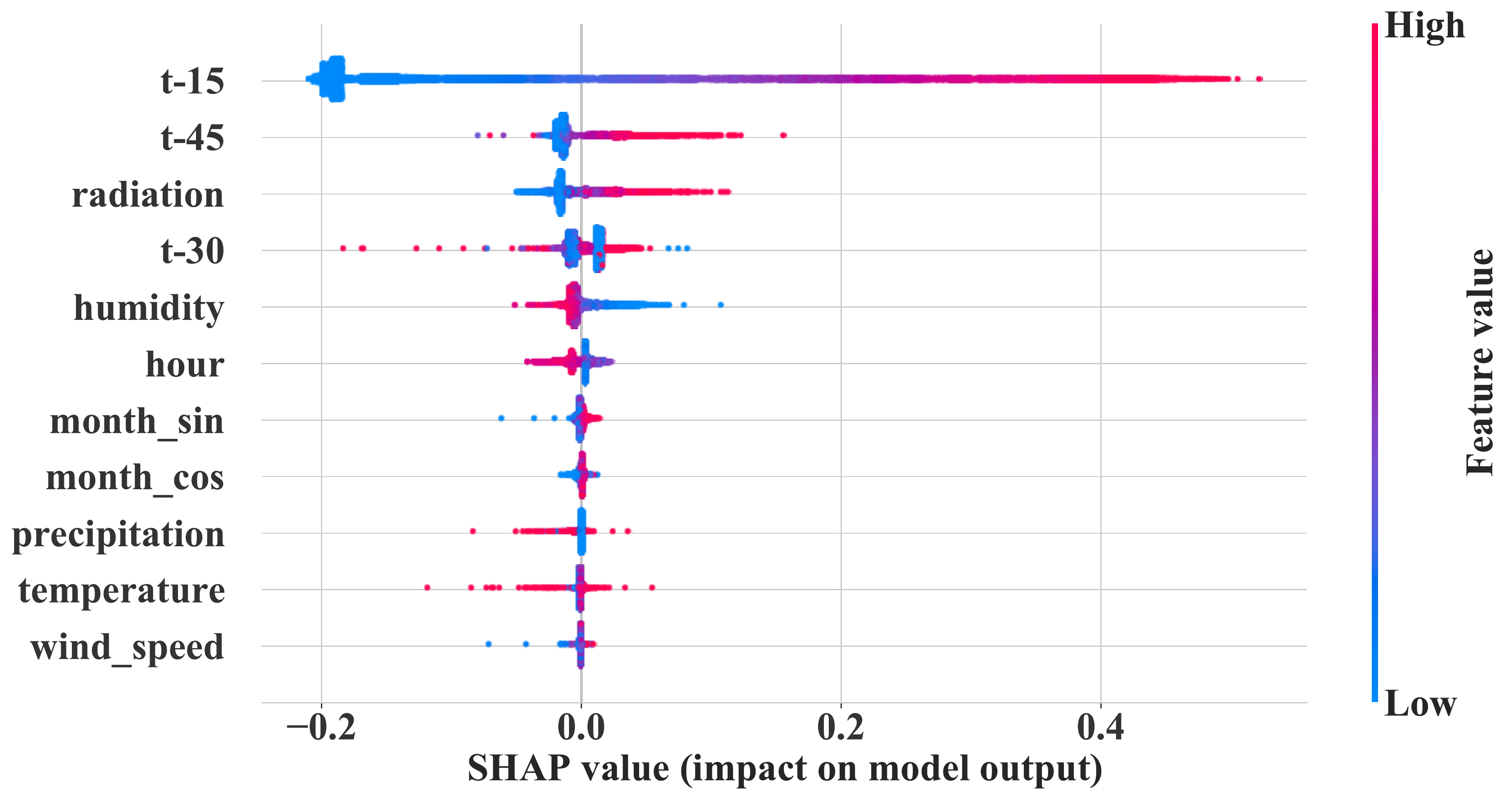}
\caption{PVP1: Summary plot of the point forecast model. Each point corresponds to an individual training example. The color of the point denotes how big of small is the value of a given feature. The horizontal position of a point indicates the influence of that feature on the model output.}
\label{fig:summary_yearly_mean_ing}
\end{figure}

However, global feature importance provides limited information about the model. Instead, we use the SHAP summary plot that summarizes the local explanations in a compact yet information-rich way, as shown in Fig.~\ref{fig:summary_yearly_mean_ing}. In this plot, each dot corresponds to an individual prediction and its position along the x-axis corresponds to the impact of the respective feature on the model output. Additionally, the color of the dot denotes the feature value in order to highlight to what extent different feature values affect the result. For instance, high t-15 values tend to push the model output to higher values, whereas close to zero t-15 values push the model output to decrease. Hence, the higher the t-15 values are, the bigger will be their impact on the model output. The vertical dispersion of the dots implies a higher number of observations with a similar impact. In this context, those SHAP summary plots reveal information about both the magnitude and the direction of each feature's impact as well as the number of observations with those characteristics. Note that all models, i.e., both for PVP1 and PVP2, yielded almost identical SHAP plots. Therefore, in order to avoid redundancy, we present the respective SHAP plots of a randomly selected model from PVP1.

Regarding the rest of the features, high radiation and t-45 values seem to have a bigger impact on the model output than their corresponding lower values. This observation may be attributed to the fact that for low radiation and lagged power values, the information about t-15 seems to suffice for the model to yield a prediction. On the contrary, for higher radiation and lagged power values, where the PV generates power, the model requires to employ more features in order to make an accurate prediction. The majority of \mbox{t-30} observations show a lower impact on the model output possibly due to its high correlation with t-15 (the model simply uses t-15) as identified in Fig.~\ref{fig:correlations}. Nonetheless, there are points where the feature t-30 manifests a big influence on the output; at those points, it is likely that t-30 interacts strongly with other features. Moreover, bigger humidity values lead to smaller power predictions and vice versa, following the physical properties of a PV cell, whose efficiency drops in high humidity \cite{mekhilef2012effect}. Similar common sense explanations apply to the time of the day feature (hour), where mid-day hours have a ``positive" impact (pushes the output to increase) on the model output since the bell-shape curve reaches its peak during that time period (Fig.~\ref{fig:power_hour}), late hours push the result towards smaller values, and early morning hours towards larger ones. As for the month, a subset of high $\text{month\_sin}$ values (corresponding to March and April) seems to have a positive effect on the predicted PV output whereas the big positive $\text{month\_cos}$ values ($=1$) have no impact on the model output. The latter may be attributed to November, December, and January (mostly winter months), where the power generation is low and thus, it is likely that the model exploits features such as t-15 and radiation to make a prediction. Finally, precipitation, temperature, and wind speed show predominantly no influence on point forecasts. Yet those features can be of high importance for a few examples, as indicated by their long tails in Fig.~\ref{fig:summary_yearly_mean_ing}.

\begin{figure}[t!] 
\includegraphics[width=\linewidth]{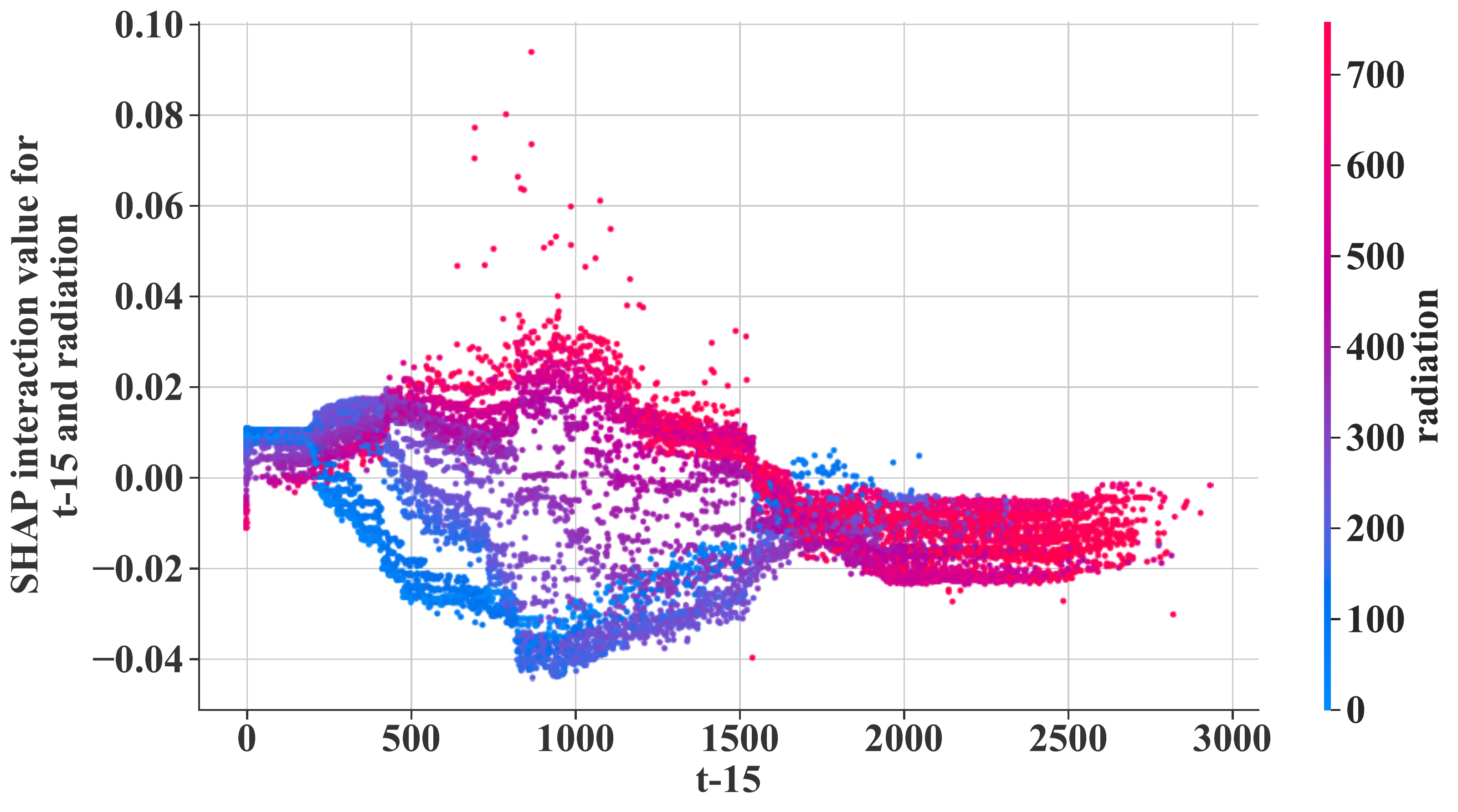}
\caption{PVP1: Interaction plots between t-15 and radiation (point forecast).}
\label{fig:t_15_radiation_mean_ing}
\end{figure}
\begin{figure}[t!] 
\includegraphics[width=\linewidth]{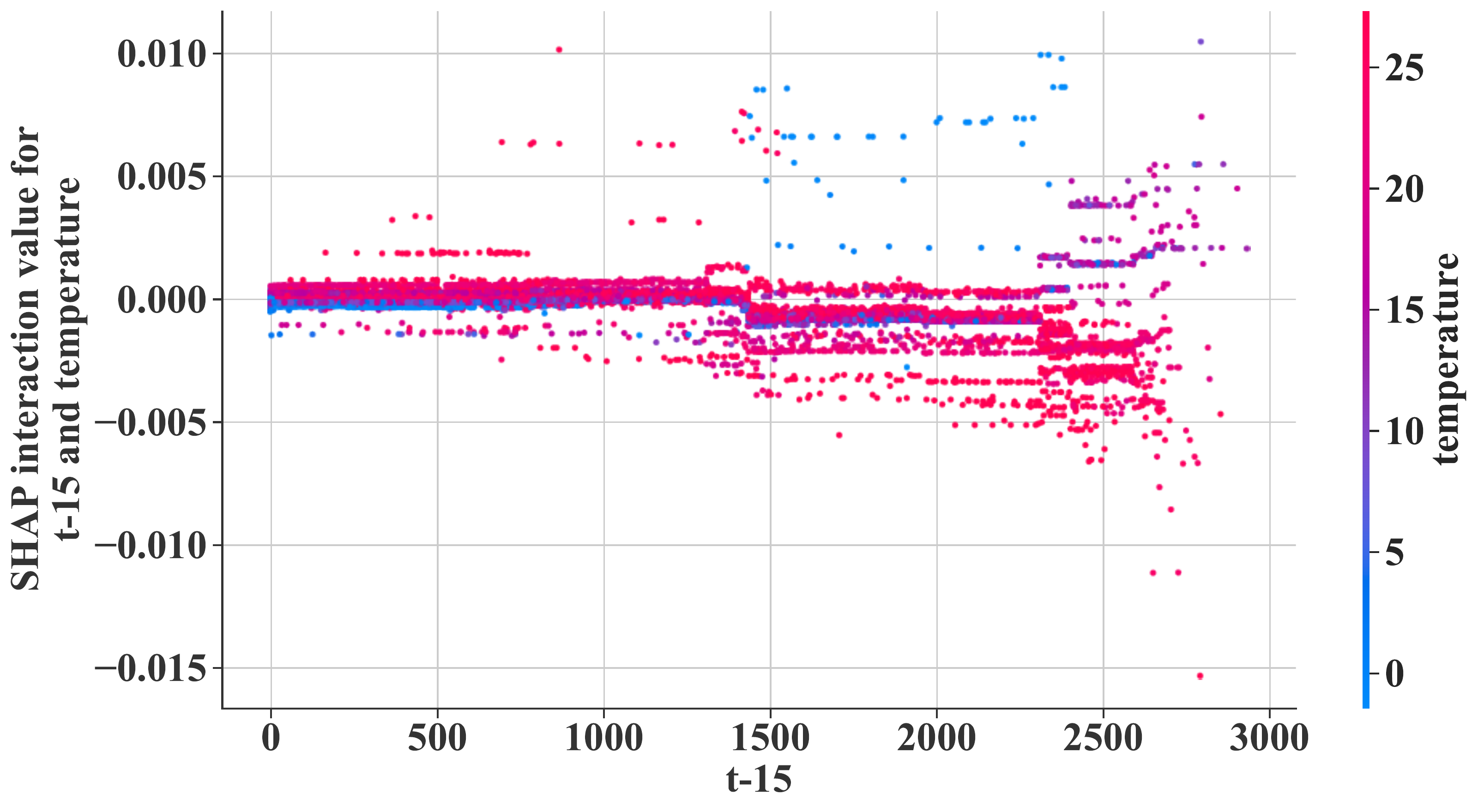}
\caption{PVP1: Interaction plots between t-15 and temperature (point forecast).}
\label{fig:t_15_temperature_mean_ing}
\end{figure}
\begin{figure}[t!] 
\includegraphics[width=\linewidth]{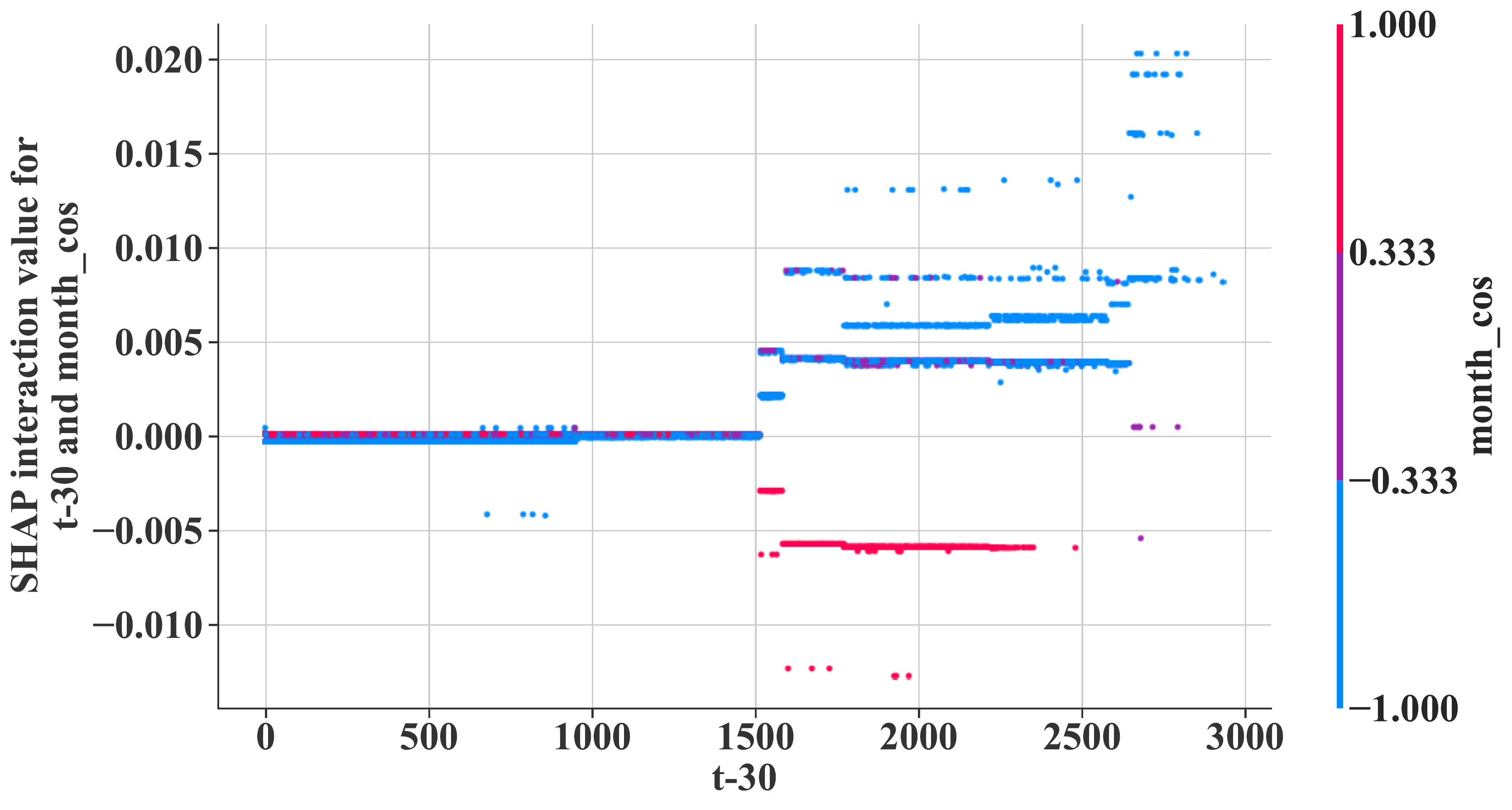}
\caption{PVP1: Interaction plots between t-30 and $\text{month\_cos}$ (point forecast).}
\label{fig:month_cos_t_30_mean_ing}
\end{figure}
\begin{figure}[t!] 
\includegraphics[width=\linewidth]{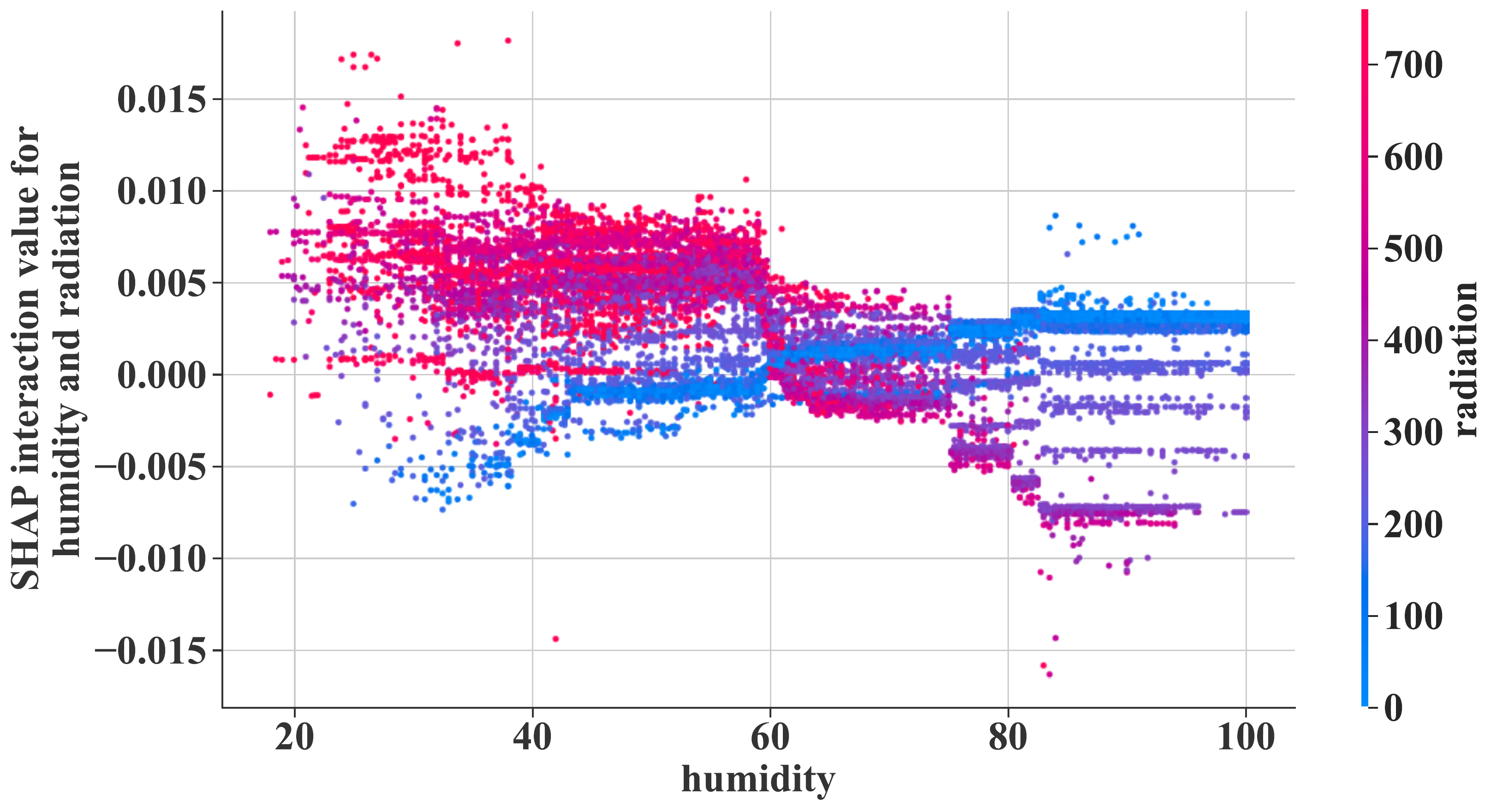}
\caption{PVP1: Interaction plots between humidity and radiation (point forecast).}
\label{fig:humidity_radiation_mean_ing}
\end{figure}

In addition to SHAP summary plots, we introduce here the derived SHAP interaction plots, which underline the complex, nonlinear feature interactions developed by the model. Importantly, high interaction values have been developed between pairs of features with high-high importance and high-low importance whereas low ranked features do not show remarkable interactions with each other. In this context, we present a set of representative SHAP interaction plots (Fig.~\ref{fig:t_15_radiation_mean_ing}-\ref{fig:humidity_radiation_mean_ing}). As derived by Fig.~\ref{fig:t_15_radiation_mean_ing}, t-15 values within the range of \SI{0.5}{MW}-\SI{1.5}{MW} can have an utterly different influence on the model predictions depending on the radiation values. As expected, high radiation leads to higher and low radiation to lower power predictions. For higher lagged power values ($>$\SI{1.5}{MW}), the interaction between t-15 and radiation tends to decrease. This can be attributed to the fact that the information about radiation is redundant when the PV park generates high power. As for the temperature, its interaction with t-15 starts at power values bigger than \SI{2}{MW}, where the model exploits also the temperature information (Fig.~\ref{fig:t_15_temperature_mean_ing}). In this region, high temperatures lead to a negative influence of the output power whereas medium temperatures within the range of 10-15\textdegree{}C have a positive impact, respectively. This can be explained by the decline of PV panel efficiency at higher temperatures \cite{coskun2017sensitivity}. In Fig.~\ref{fig:month_cos_t_30_mean_ing}, we observe how the model combines the information about the month with the power of t-30. During the spring and summer months ($\text{month\_cos} < 0$) and for power values bigger than \SI{1.5}{MW}, the model will tend to yield higher power predictions whereas during the autumn and winter months ($\text{month\_cos} > 0$), the model will generate lower power predictions for the same power range. In the interaction plots of Fig.~\ref{fig:humidity_radiation_mean_ing}, the combination of humidity and radiation shows a significant influence on the model output. For high humidity values, an increase in radiation pushes the generation to decrease whereas for low humidity values, the predicted PV generation increases as the radiation increases. Overall, the gradient boosting trees modeling the point forecasts (mean of the distribution) show to be consistent with common sense and physical properties characterizing the PV generation.

Note that the SHAP values describe the relationships that the model has learned and not the linear correlation between features. In this regard, the feature importance derived by the SHAP values may significantly differ from the linear correlation coefficients, as highly nonlinear interactions were discovered. Only in a linear model, the feature importance would probably coincide with the correlation coefficients.

\begin{figure}[t!] 
\includegraphics[width=\linewidth]{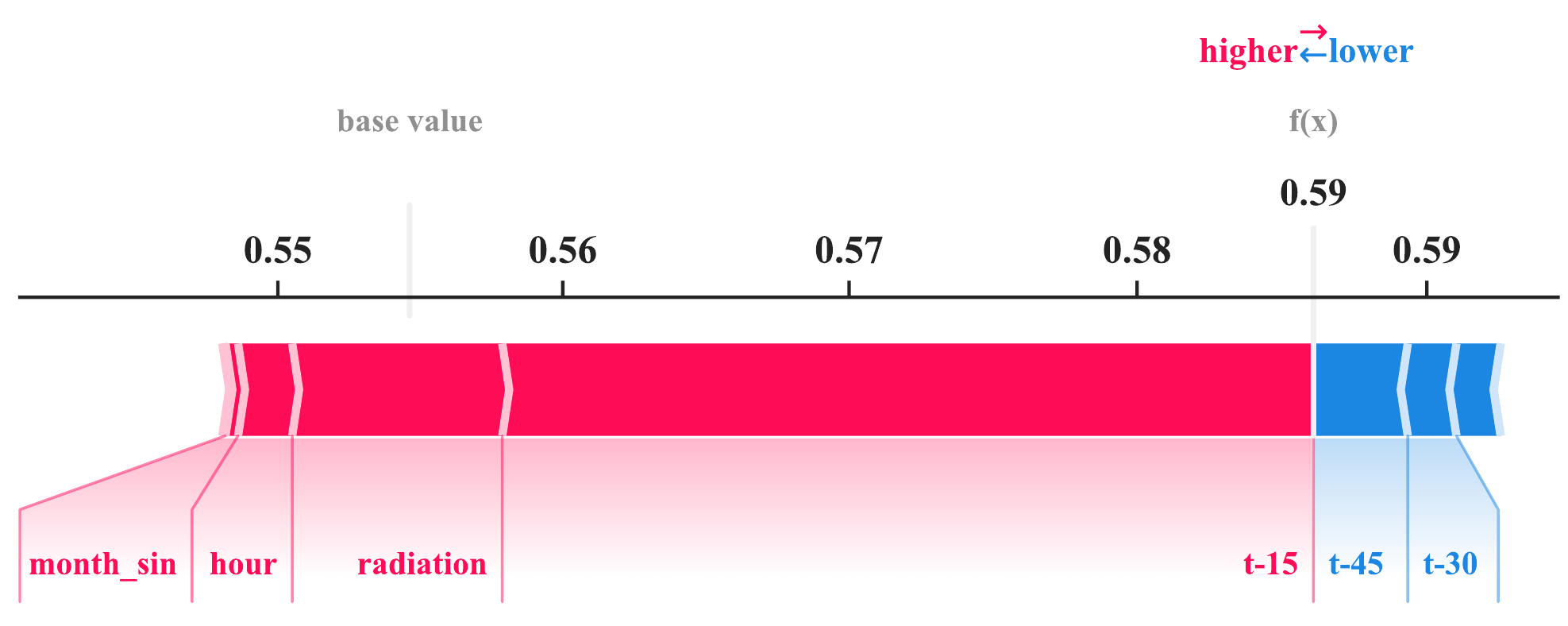}
\caption{PVP1: Force plot showing the contribution of each feature for a random point forecast.}
\label{fig:individual_mean_ing_851}
\end{figure}

Finally, a representative example of a ``force" plot corresponding to an individual prediction is presented in Fig.~\ref{fig:individual_mean_ing_851}. A force plot visually describes the magnitude and direction of the influence that each feature has on the output of an individual prediction. It is worth pointing out that the scope of force plots is limited to interpreting individual predictions whereas no general inference can be derived about the model behavior. In this example, t-15 = \SI{0.89}{MW}, t-30 = \SI{0.56}{MW}, t-15 = \SI{0.54}{MW}, radiation = \SI{515.2}{W/m^2}, hour = 10:45, and the output corresponds to the normalized one (0.59). In this regard, morning hours, high radiation and moderate t-15 values will push the model output to increase, whereas the low past t-30 and t-45 will have the opposite effect but with a lower magnitude. Note that the features with negligible contribution to the final output are not displayed.

\subsection{Interpretation of model uncertainty}

\begin{figure}[t!] 
\includegraphics[width=\linewidth]{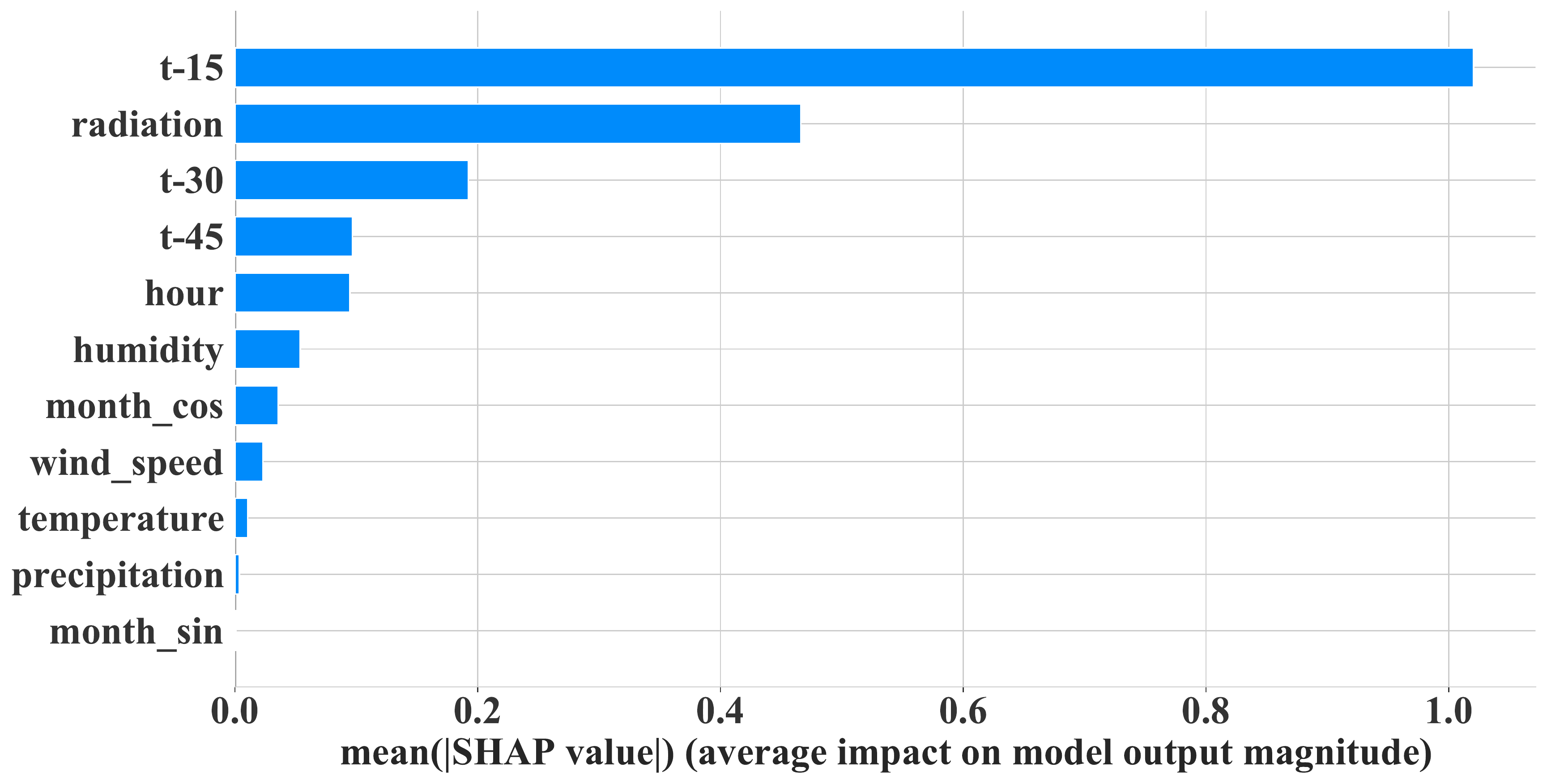}
\caption{PVP1: Global feature importance of the model uncertainty. Each bar plot correspond to the global importance of an individual feature.}
\label{fig:bar_yearly_std_ing}
\end{figure}

\begin{figure}[t!] 
\includegraphics[width=\linewidth]{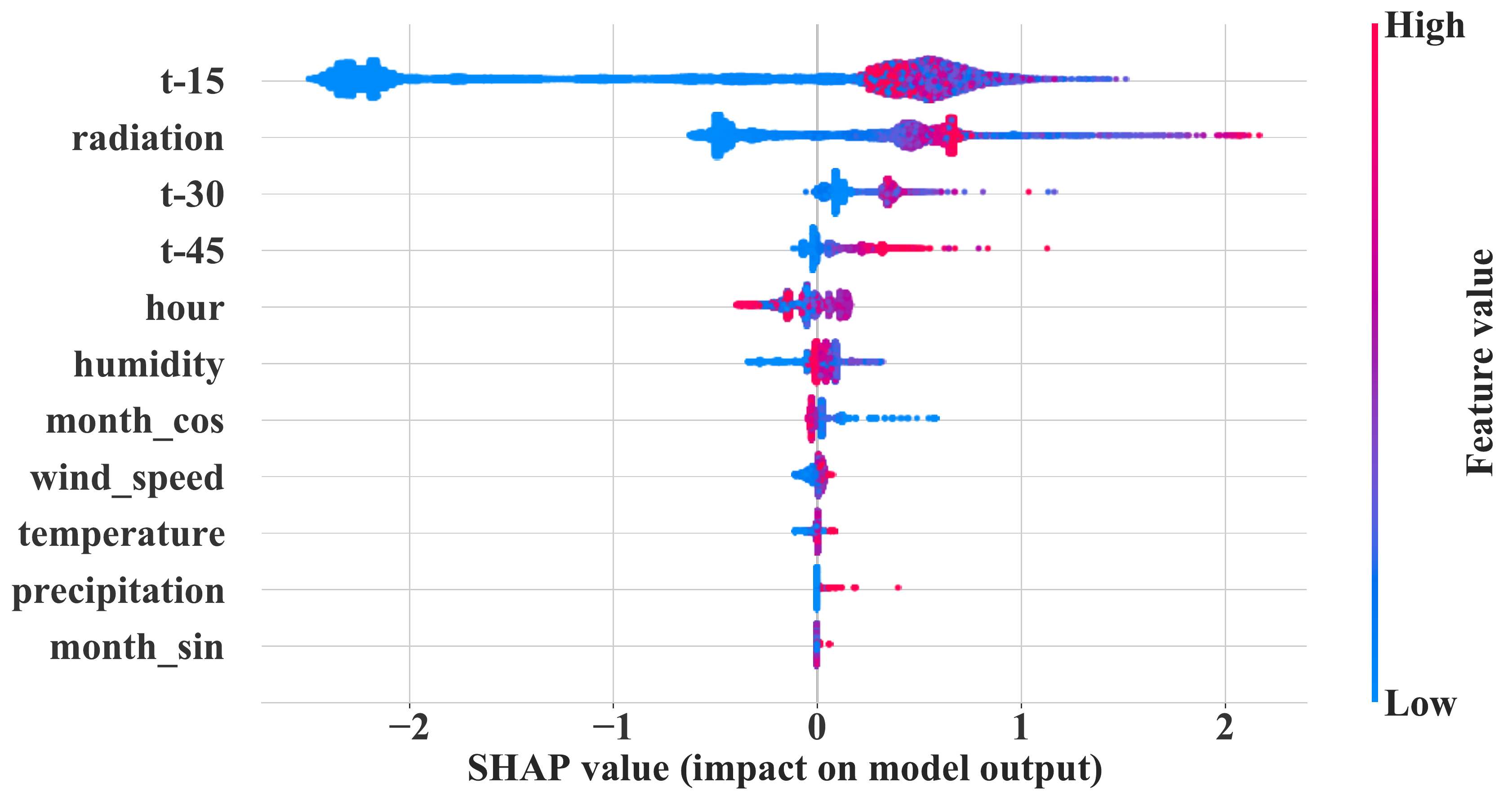}
\caption{PVP1: Summary plot of the model uncertainty. Each point corresponds to an individual training example. The color of the point denotes how big of small is the value of a given feature. The horizontal position of a point indicates the influence of that feature on the model output.}
\label{fig:summary_yearly_std_ing}
\end{figure}

One of the main contributions of this work is to analyze and understand how the model adjusts its predictive uncertainty. To do so, we perform the same analysis as in the previous section but for the PIs instead of the point forecasts. Starting with global explanation values, Fig.~\ref{fig:bar_yearly_std_ing} and \ref{fig:summary_yearly_std_ing} illustrate the global feature importance using bar plots and SHAP summary plots, respectively. In general, the order of the feature importance follows mostly the same pattern as in the point forecast model. Nevertheless, the influence of t-15 on PI estimation compared to the rest of the features has been reduced while the latter have gained importance. Only $\text{month\_sin}$ seems to have no influence on the model output, which may be explained by the fact that $\text{month\_cos}$ splits the unit circle into spring-summer (higher power output) and autumn-winter (lower power output) semicircles and thus, $\text{month\_cos}$ reflects all necessary seasonal information.

Regarding the summary plots, the uncertainty of the model decreases significantly for small t-15 values. For instance, if the PV park does not generate power, it is very likely that it will continue doing so in the next 15 min. For higher power values, t-15 has a comparable impact for most of the observations indicating the stochastic nature of PV generation. Consequently, higher radiation usually means lower confidence about the model prediction. This behavior can be also observed in the t-30 and t-45 features. The only difference between them is that low two past power values have almost no effect on the model output. In this regard, t-15 and radiation seem to suffice for the model to increase its confidence. Furthermore, mid-day hours, where the power generation is high, seem to have an negative effect on the model's confidence whereas night hours seem to be modeled with higher confidence, like the morning hours. This can be readily explained by the high variance of power values during mid-day hours (Fig.~\ref{fig:power_hour}). Low humidity values seem to have a determinant role in reducing the uncertainty of the output, while high humidity values do not contribute to model confidence. Medium humidity values have a small impact on the uncertainty, possibly due to the high variance of the respective humidity data. As mentioned, low $\text{month\_cos}$ values correspond to spring and summer, where the generation is high and thus, it is subject to many variations. Therefore the model is less confident during those two seasons and more confident during autumn and winter where generation is not expected to be high. As in the point forecast model, temperature and wind speed do not have a remarkable impact on the result while precipitation has almost zero influence on model uncertainty. Yet a limited number of high precipitation points have a negative impact on model's confidence. Therefore, the overall contribution of those three features may be questioned and will be studied in the next section. 

\begin{figure}[t!] 
\includegraphics[width=\linewidth]{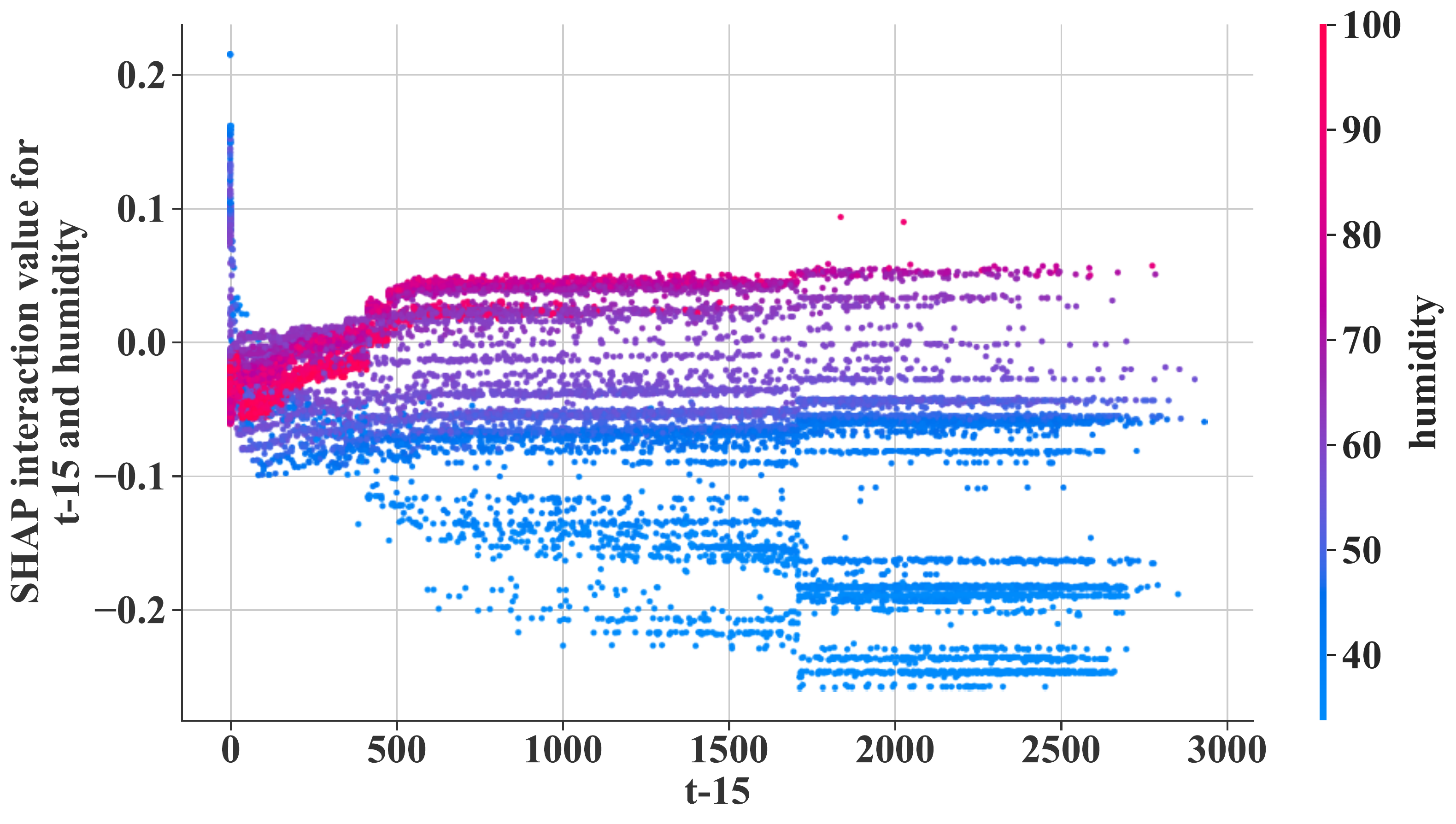}
\caption{PVP1: Interaction plots between t-15 and humidity (PI forecast).}
\label{fig:t_15_humidity_std_ing}
\end{figure}
\begin{figure}[t!] 
\includegraphics[width=\linewidth]{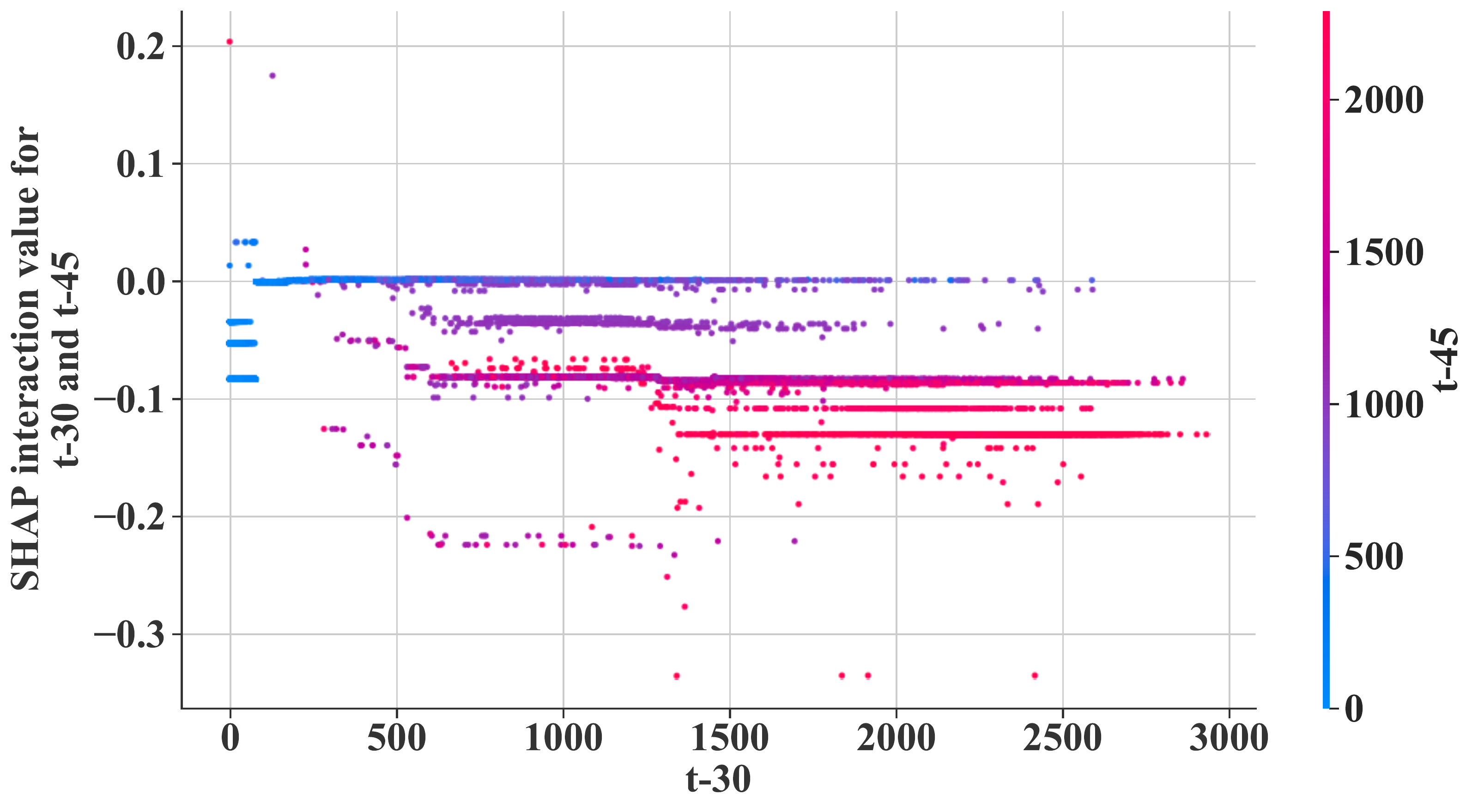}
\caption{PVP1: Interaction plots between t-30 and t-45 (PI forecast).}
\label{fig:t_30_t_45_std_ing}
\end{figure}
\begin{figure}[t!] 
\includegraphics[width=\linewidth]{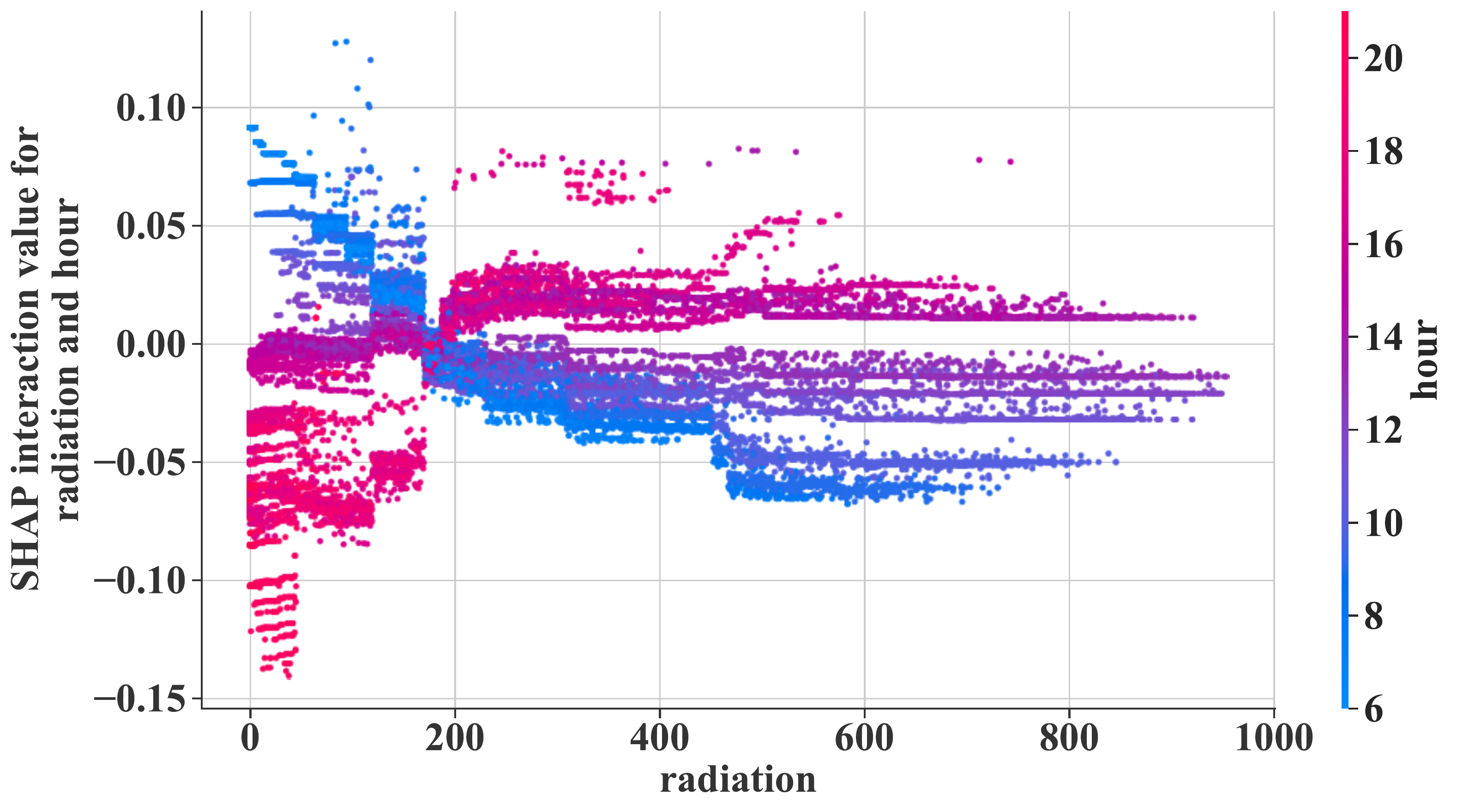}
\caption{PVP1: Interaction plots between radiation and hour (PI forecast).}
\label{fig:radiation_hour_std_ing}
\end{figure}
\begin{figure}[t!] 
\includegraphics[width=\linewidth]{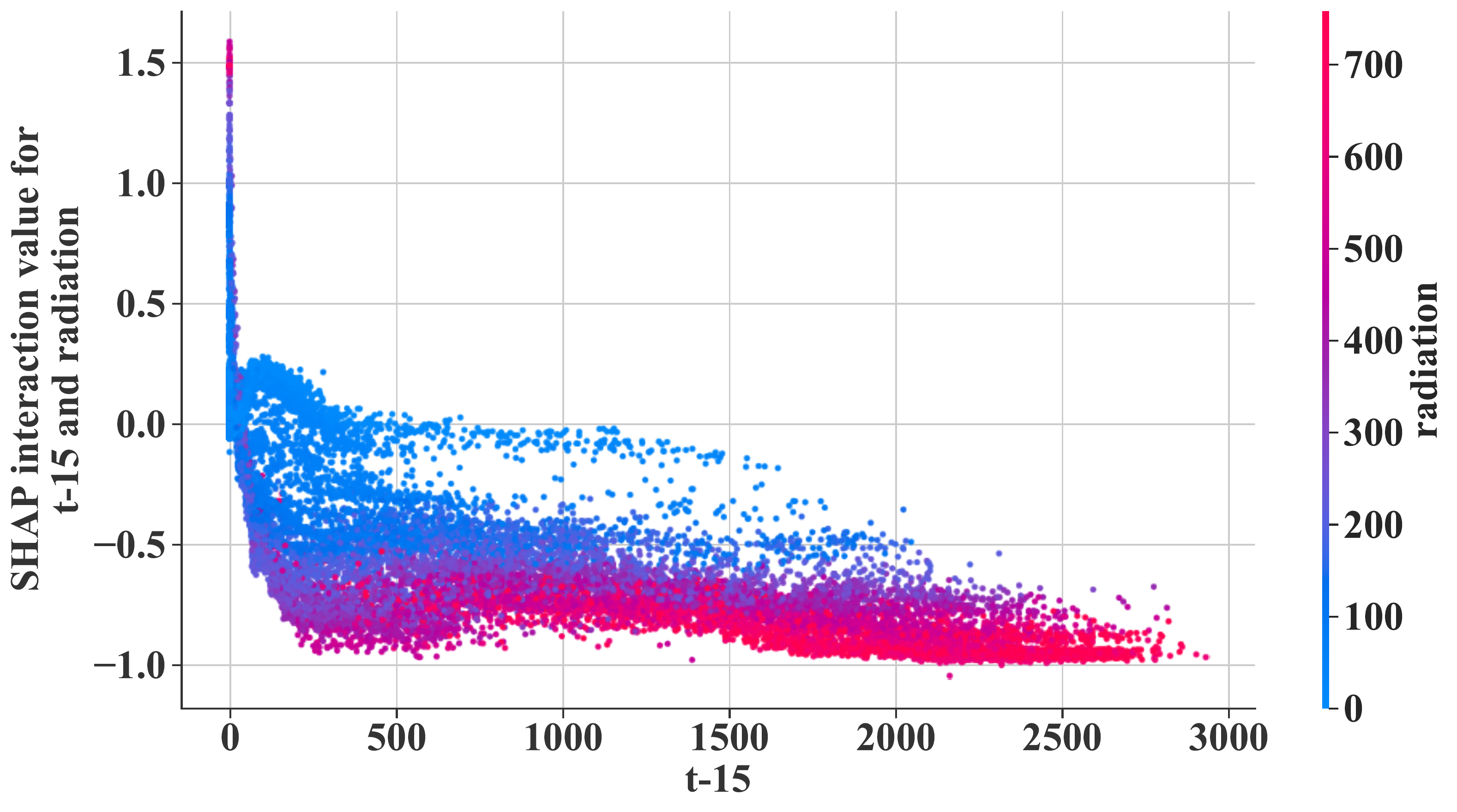}
\caption{PVP1: Interaction plots between t-15 and radiation (PI forecast).}
\label{fig:t_15_radiation_std_ing}
\end{figure}

A representative set of SHAP interaction plots corresponding to how the model handles uncertainty by combining different feature pairs is shown in Fig.~\ref{fig:t_15_humidity_std_ing}-\ref{fig:t_15_radiation_std_ing}. Similar to the interaction plots of point forecasts, high interaction values have been developed between pairs of features with high-high importance and high-low importance. An interesting observation is that the model exploits the humidity information in order to increase its confidence about a prediction, as depicted in Fig.~\ref{fig:t_15_humidity_std_ing}. Low humidity values have a significant positive effect, whereas the rest of humidity values are not considered in the estimation of the PIs. Moreover, similar past power values have only a positive impact on the model's confidence. The bigger the power values are, the bigger becomes the confidence about the model's prediction (Fig.~\ref{fig:t_30_t_45_std_ing}). As for the influence of the time of the day (hour), Fig.~\ref{fig:radiation_hour_std_ing} highlights the nonlinear interactions between the radiation and hour. For low radiation, the time of the day plays a crucial role in model's uncertainty. Morning hours are characterized by an increase in uncertainty whereas low radiance in night hours has a clear positive impact since no big variations in power generation are expected compared to the morning hours. For radiation values above \SI{200}{W/m^2}, the interaction between those features decreases. Yet early morning hours combined with high radiation tend to decrease the predictive uncertainty. Finally, as in point forecasts, radiation and t-15 are also strongly connected in the uncertainty estimation. Radiation seems to provide an extra confidence in the model, when it is combined with the t-15 information, as confirmed by Fig.~\ref{fig:t_15_radiation_std_ing}. Based on the aforementioned analysis, the gradient boosting trees estimating the standard deviation (uncertainty) of the predictive distribution manifests that the model can successfully encode the stochastic nature of PV generation while following physical properties and common sense.

\begin{figure}[b!] 
\includegraphics[width=\linewidth]{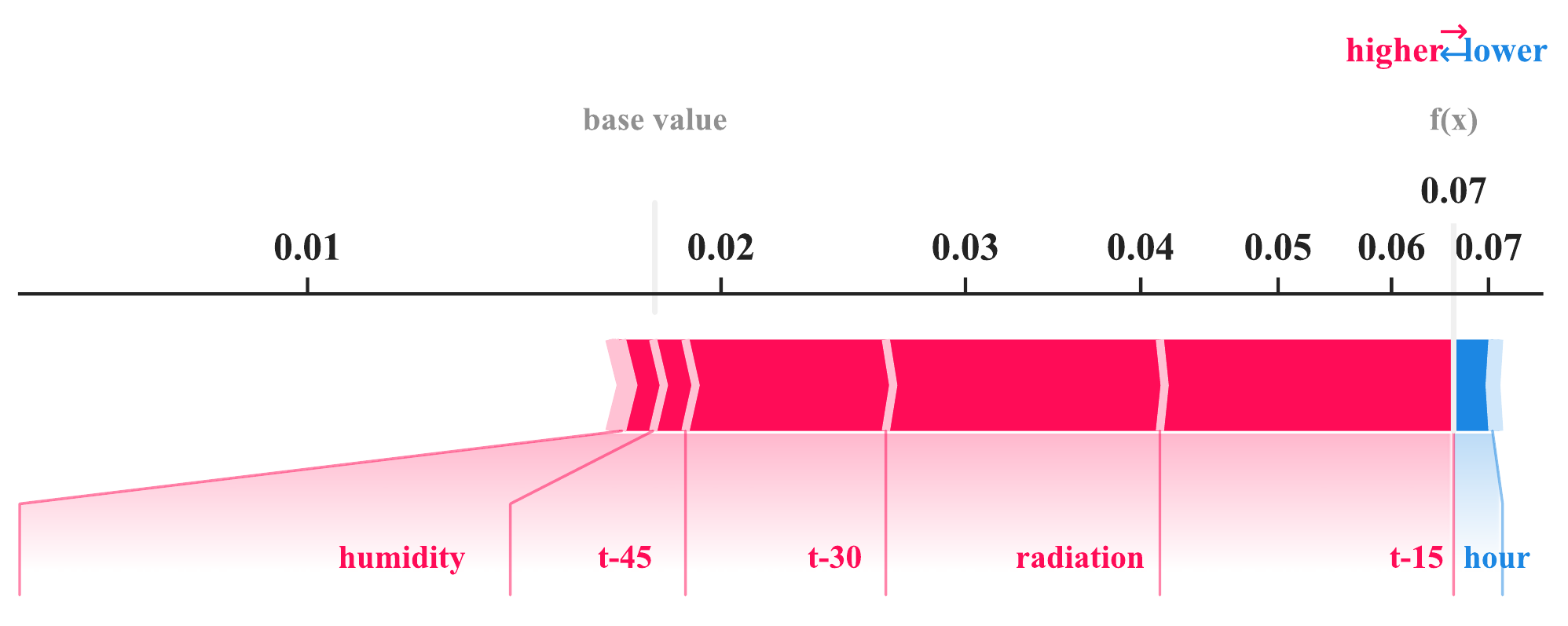}
\caption{PVP1: Force plot showing the contribution of each feature to the uncertainty estimation for a random prediction.}
\label{fig:individual_std_ing_851}
\end{figure}

Using the same sample as in Fig.~\ref{fig:individual_mean_ing_851}, we illustrate the corresponding force plot with respect to the uncertainty estimation in Fig.~\ref{fig:individual_std_ing_851}. Compared to the point forecast, this model exploits predominantly the information of the same features in order to estimate the PIs. In particular, past power values, radiation, and humidity contribute to an increase in the model's uncertainty (standard deviation), while the time of day tends to influence positively the model output (reducing uncertainty).

\subsection{Model performance with a subset of features}
Based on the previous analysis, we identified that the weather information of precipitation, temperature, and wind speed are deployed only for a small number of observations by the point forecasting model while their contribution to estimating the uncertainty related to a prediction is rather negative. Therefore, we discarded those features from our datasets and we retrained the models. In Table \ref{tab:features}, we present the average forecasting metrics yielded by the test sets for the case of using all the aforementioned features and for the case, in which precipitation, temperature, and wind speed are discarded. In particular, there is an increase in accuracy of around 6\% for RMSE and around 10\% for CRPS. This result may be caused by a local optimum solution in the training of the model where all features were employed due to the higher dimensionality of the problem. 

Therefore, following the proposed methodology, one can fully understand how a trained model works, determine the features leading to an increase in uncertainty, and possibly discard them. It is worth pointing out that there may be specific examples where those features are useful for the model and lead to lower prediction errors. In the context of the overall performance however, they may have a rather negative impact.

\begin{table}[b]
\caption{Averagae forecasting metrics using a training set with all features and a reduced one.}
\label{tab:features}
\begin{tabular}{lcc}
\hline
           & \textbf{NGBoost (all features)} & \textbf{NGBoost (reduced features)} \\ \hline
MAE (pu)   & 0.0367                          & 0.0325                              \\
RMSE (pu)  & 0.0617                          & 0.0577                              \\
MBE (pu)   & -0.0007                         & -0.0002                             \\
PICP (-)   & 0.8882                          & 0.889                              \\
PINAW (pu) & 0.1732                          & 0.1805                              \\
CRPS (pu)  & 0.0274                          & 0.0245                              \\ \hline
\end{tabular}
\end{table}

\section{Discussion}
The optimal PV integration is a challenging problem due to volatile weather conditions leading to intermittent generation. The latter has a direct impact on system stability and reliable electricity supply and thus, accurate and reliable PV forecasting techniques are of utmost importance. In this context, a small number of probabilistic models have been recently introduced in the literature aiming at providing more information about the derived forecasts compared to deterministic methods. This additional information corresponds to the uncertainty of the predictions and is usually expressed in the form of PIs. However, most of the existing probabilistic approaches are complex black box models characterized by high training requirements and extensive hyperparameter tuning. Furthermore, the lack of interpretability may prevent their practical implementation in a safety critical system as the power system.

As a solution, we propose an open source probabilistic forecasting method that minimizes the expert knowledge required, does not require extensive training times, yields highly accurate and reliable forecasts, and finally, provides full transparency on its predictions. To do so, we propose a two stage forecasting framework. In the first stage, we leverage the newly introduced NGBoost algorithm in order to yield accurate and reliable probabilistic forecasts while in the second stage, we calculate the SHAP values for both the point forecasts as well as the derived PIs.

Based on a thorough comparison using both deterministic and probabilistic metrics, we showed that  NGBoost can achieve better performance than GP and LUBE, regardless of the seasonal weather and power variations. Importantly, NGBoost can be trained more than ten times faster than the other two algorithms. This can be considered as a great advantage for a machine learning approach since most of those approaches may require many trial and error attempts for finding the best combination of features and hyperparameters. It is also worth mentioning that the combination of gradient boosting trees of depth 3, a learning rate of 0.01, 500 boosting iteration, a log score, and a Gaussian distribution has been empirically found to be a suitable choice for the NGBoost hyperparameters in the context of PV power forecasting. Those parameters could serve as a starting point the practical implementation of the proposed method in a real-world scenario.

It worth pointing out that an alternative would be to use gradient boosting or even extreme gradient boosting (XGBoost) for point forecast and then quantile regression or bootstrapping for the prediction intervals. Nevertheless, this would require the training and the coding of two models and thus, it would  increase the effort for the practical implementation. As we want to minimize the effort and the expert knowledge required while maintaining high accuracy levels, we have opted for a single model that can yield both point forecasts and prediction intervals.

Regarding the second stage, we demonstrated that local feature importance is particularly advantageous with respect to interpreting the output of a model. By using the theoretically optimal SHAP values as feature attributions, we are able to determine the magnitude and the direction of the influence of each individual feature on the final result. Specifically, the SHAP method was applied for the first time on both the prediction point and the PI models. A detailed analysis of the derived SHAP values revealed that the forecasting models came up with some nonlinear feature relationships that follow known physical properties and human logic and intuition. This outcome may have a significant impact on tackling the missing trust in machine learning models and thus, help them become widespread.  

Furthermore, the optimal feature selection is a rather challenging task and it is highly dependent on the deployed models and the given dataset \cite{ahmed2020review}. More features usually lead to higher complexity and longer training times. Against this background, understanding the influence of each feature on the model may alleviate the problem of feature selection. In this context, features with a very limited contribution were discarded and the model was retrained. As a result, the new model achieved higher accuracy and sharper probabilistic forecasts. This outcome may be justified by the fact that the dimensionality of the parameter estimation problem increases with the number of features. Therefore, the iterative optimization may yield a suboptimal solution due to the higher dimensionality.  

\section{Conclusion}

The main findings and gains of the proposed transparent model can be summarized in the following points:
\begin{itemize}
\item It was observed that no counter-intuitive or surprising relationship was developed by the proposed model. This finding is of utmost importance considering that debugging machine learning models is an extremely challenging task. Even at big technology companies, many bugs in machine learning pipelines may not be discovered \cite{zinkevich2017rules}. Once a model is trained, no insight is given about whether it has learned meaningful relationships and the test set usually comprises a rather small amount of data. Therefore, model weaknesses or unrealistic learned relationships may not be detected. In this context, the proposed model seems to not have any evident bugs or biases.  
\item An increase of around 6\% in RMSE and 10\% in CRPS is achieved by identifying and then applying the most important available features. This is one of the very few cases in which SHAP values are actually used for increasing a model's performance and not just explaining the model's predictions.
\item The proposed model can be employed by the various stakeholders, such as system operators or traders, whose decisions require high transparency and contain financial risks. As a general rule, they are often reluctant in deploying black box models in practice. To this end, the proposed approach can exploit the full potential of a complex machine learning model while providing full transparency on its predictions.
\end{itemize}

As a next step, a detailed analysis of the feature values yielding large prediction errors is to be performed. We expect that this will reveal important information about when and under which conditions bigger forecasting errors are observed. Future work will also aim at the implementation of the proposed method in other forecasting problems related to power systems, e.g., wind power forecasting, load forecasting, etc., in order to identify whether the derived models are also built based on physical properties and human intuition.

\section{Acknowledgements}
The authors would like to thank NetzeBW GmbH (DSO) for providing the data for this research.

\ifCLASSOPTIONcaptionsoff
  \newpage
\fi



\bibliographystyle{IEEEtran}
\bibliography{IEEEabrv, mitre_ngboost}

\end{document}